\pdfoutput=1
\documentclass[12pt]{article}
\usepackage[utf8]{inputenc}
\usepackage{cite}
\usepackage{hyperref}
\hypersetup{colorlinks=true,linkcolor=blue2,citecolor=red2,urlcolor=green2,pdfencoding=auto,linktocpage}
\usepackage{amsmath,amssymb,amsbsy,amstext,amsthm,simplewick,amsfonts}
\usepackage{mathrsfs}
\usepackage{graphicx}
\usepackage{wrapfig}
\usepackage{upgreek}
\usepackage{bm} %-> bold math
\usepackage{framed}
\usepackage{bbm}
\usepackage{textcomp}
\usepackage{adjustbox}
\usepackage{makecell}
\usepackage{tcolorbox}
\usepackage{empheq}
\usepackage[normalem]{ulem}
\usepackage{enumitem}
\usepackage{braket} 
\usepackage{array}
\usepackage{dsfont}
\usepackage{physics}
\usepackage{ulem}
\usepackage{xcolor}
\usepackage{caption}
\usepackage{subcaption}
\usepackage{tocloft}
\usepackage{tikz}
\usepackage{slashed}
\usepackage{bbm}
\usepackage{lmodern}
\usepackage{booktabs} % For professional-quality tables
\usepackage{amsmath}  % For math formatting
\usepackage{xcolor}

\usetikzlibrary{decorations.pathmorphing}

\usepackage[framemethod=default]{mdframed}

\usepackage{float}

\usepackage{tikz}
\usepackage{tikz-feynman}
\tikzfeynmanset{compat=1.1.0}
\usetikzlibrary{decorations.pathmorphing, positioning}

\newmdenv[backgroundcolor=gray!15,%
skipabove=5pt,%
skipbelow=5pt,%
leftmargin=2pt,%
rightmargin=2pt,%
innertopmargin=-6pt,%
innerbottommargin=5pt,%
innerleftmargin=5pt,%
innerrightmargin=5pt,%
splittopskip=0pt,%
splitbottomskip=0pt,%
linewidth=0pt,%
nobreak=true]%
{keyeqn}

\newmdenv[backgroundcolor=gray!15,%
skipabove=5pt,%
skipbelow=5pt,%
leftmargin=2pt,%
rightmargin=2pt,%
innertopmargin=-2pt,%
innerbottommargin=5pt,%
innerleftmargin=5pt,%
innerrightmargin=5pt,%
splittopskip=0pt,%
splitbottomskip=0pt,%
linewidth=0pt,%
nobreak=true]%
{keythrm}

\graphicspath{{Fig/}}

\definecolor{red2}{RGB}{214, 39, 40}
\definecolor{green2}{RGB}{0,170,0}
\definecolor{blue2}{RGB}{0,100,200}
\definecolor{magenta2}{RGB}{191,64,191}
\definecolor{purple2}{RGB}{112,48,160}
\definecolor{orange2}{RGB}{255,192,0}

\def\bgm{\begin{matrix}}
\def\edm{\end{matrix}}

\numberwithin{equation}{section}

\allowdisplaybreaks[1]
\graphicspath{{Fig/}}
\begin{document}

\begin{titlepage}
\centering

\vspace*{-0.5cm}

{\fontsize{19}{25}\selectfont\bfseries
Primordial Correlators from a\\[3pt]
Kaluza--Klein Graviton Continuum
\par}

\vspace{0.65cm}

{\large Shuntaro Aoki\par}

\vspace{0.18cm}

{\footnotesize
\href{mailto:shuntaro.aoki@riken.jp}{shuntaro.aoki@riken.jp}
\par}

\vspace{0.4cm}

{\small\itshape
RIKEN Center for Interdisciplinary Theoretical and Mathematical Sciences
(iTHEMS),\\
Wako, Saitama 351-0198, Japan
\par}

\vspace{0.9cm}

\noindent\rule{0.86\textwidth}{0.4pt}

\vspace{0.55cm}

{\bfseries Abstract\par}

\vspace{0.45cm}

\begin{minipage}{0.86\textwidth}
\small
Cosmological collider signals are usually discussed for isolated massive
particles, whose exchange produces characteristic logarithmic oscillations in
primordial correlators.  In this work, we study how this signal is modified
when the exchanged states form a continuous mass spectrum.  We first develop a
spectral representation for inflationary correlators mediated by a continuum
field.  In the soft limit, the non-analytic part of the seed function is
expressed as a Fourier--Laplace transform of the spectral weight with respect
to logarithmic momentum variables.  This representation shows that a continuum
superposes clock signals with different frequencies and can dephase the sharp
oscillation associated with a single massive particle.

We then realize this mechanism in an RS2-like inflationary braneworld, where
the inflaton is localized on a de Sitter brane in a five-dimensional AdS bulk.
The tensor sector contains a localized massless graviton and a continuum of
Kaluza--Klein gravitons starting at \(m=3H/2\).  We derive the KK wavefunctions
and identify the brane spectral weight fixed by the continuum wavefunction on
the brane.  Applying this weight to the inflaton four-point function, we find a
smooth seed function rather than a persistent logarithmic clock oscillation.
This behavior follows from the fact that the continuum starts at zero clock
frequency, \(\mu=0\), while the KK spectral weight vanishes near threshold.  Our
results provide a concrete higher-dimensional example in which deviations from
the standard cosmological collider signal encode the structure of a continuum
high-energy spectrum.
\end{minipage}

\vspace{0.7cm}

\noindent\rule{0.86\textwidth}{0.4pt}

\end{titlepage}

%\restoregeometry

\newpage
\setcounter{page}{2}
{
	\tableofcontents
}

\newpage
%\newpage

%%%%%%%%%%%%%%%%%%%%%%%%%%%%%%%%%%%%%%%%%%%

%%%%%%%%%%%%%%%%%%%%%%%%%%%%%%%%%%%%%

\section{Introduction}

Cosmological collider physics provides a way to probe massive particles during
inflation through their imprints on primordial correlation functions
\cite{Chen:2009zp,Baumann:2011nk,Noumi:2012vr,Arkani-Hamed:2015bza}.  A massive particle with
mass of order the Hubble scale can be produced during inflation and subsequently
decay into inflaton fluctuations.  In momentum space, this process gives rise to
a characteristic non-analytic dependence of primordial correlators.  In
particular, for a particle in the principal series, the signal oscillates as a
function of logarithmic momentum ratios, with a frequency determined by the
mass of the exchanged particle.  This ``clock signal'' is the standard template
for identifying the mass spectrum of heavy particles during inflation.

In recent years, significant progress has been made in computing such
cosmological correlators.  A variety of methods, including
differential-equation techniques
\cite{Arkani-Hamed:2018kmz,Baumann:2019oyu,Pimentel:2022fsc,Jazayeri:2022kjy,
Qin:2023ejc,Aoki:2024uyi,Aoki:2023wdc,Liu:2024str,Qin:2025xct},
Mellin--Barnes representations
\cite{Qin:2022lva,Qin:2022fbv,Qin:2024gtr,Aoki:2026vbc},
dispersive methods
\cite{Liu:2024xyi,Chowdhury:2026upp,Das:2026vfv},
spectral decompositions
\cite{Aoki:2026vbc,Xianyu:2022jwk,Loparco:2023rug,Werth:2024mjg,Zhang:2025nzd}, KLF-space methods~\cite{
Belrhali:2026DeSitterMomentum,
Belrhali:2026KLF,
Belrhali:2026DoubleExchange
},
Laplace-space representations~\cite{
Belrhali:2026Laplace,
Belrhali:2026MassiveLaplace
},
and cutting-rule approaches
\cite{Tong:2021wai,Qin:2023bjk,Ema:2024hkj},
have led to increasingly explicit analytic expressions for cosmological
correlators mediated by massive particles. As a result, the standard
cosmological-collider signal from an isolated massive particle is now under
relatively good analytic control. Recent analytic progress has also extended to strongly mixed multifield
systems, including exact linear solutions~\cite{Huenupi:2026abj} and nonperturbative treatments
of cosmological-collider bispectra
\cite{Huenupi:2026aqc,Wang:2026lff,Pinol:2026xnl,Pinol:2026hey,Belrhali:2026uxn}. These analytic developments have also
motivated data-oriented templates for searches of cosmological-collider signals
in cosmological observations~\cite{Cabass:2024wob,Goldstein:2024bky,Sohn:2024xzd,Goldstein:2025eyj,Anbajagane:2025uro, Suman:2025tpv, Green:2026yev, Kumar:2026ogn, Philcox:2026bfa, Kumar:2026dih,Cassem:2026ygh,Philcox:2026rpn,Philcox:2026tjj}.

This progress makes it timely to ask how robust the standard signal is, and
what kinds of deviations from it can arise in more general high-energy
theories.  Such deviations are important: they may encode information not only
about the mass and spin of individual particles, but also about the structure
of the underlying spectrum and the UV completion of the inflationary sector.
Several studies have explored this direction, including interference among
multiple particles
\cite{Aoki:2020zbj,Pinol:2021aun,Aoki:2024uyi},
scale-dependent signals
\cite{Reece:2022soh,Aoki:2023wdc},
CP-violating phases
\cite{Cui:2021iie},
non-Bunch--Davies initial states
\cite{Yin:2023jlv},
ghost-mediated signals
\cite{Aoki:2025uff}, fermion-condensation effects
\cite{Tong:2023krn}, classical resonance~\cite{Chen:2022vzh,Wang:2025qww,Werth:2023CosmologicalFlow,
Pinol:2023CosmologicalFlow}, alternative to inflation~\cite{Quintin:2024boj},
and compact scalar sectors
\cite{Chakraborty:2023eoq,Chakraborty:2025myb,Chakraborty:2025mhh}.%\SA{sound speed?ghost?}

In this work, we focus on one simple but physically motivated deviation from the
standard picture: a {\it{continuous}} mass spectrum.  Instead of assuming that the
intermediate particle is an isolated state with a fixed mass, we consider a
field whose propagator is described by a spectral density over a continuum of
masses.  From the four-dimensional point of view, the corresponding
cosmological-collider signal is obtained by averaging the standard fixed-mass
signal over the continuous spectrum.  Such continuum spectra have been
discussed in particle-physics phenomenology, especially in connection with
dark-sector and continuum dark-matter models
\cite{Csaki:2021gfm,Csaki:2021xpy,Csaki:2022lnq,Ferrante:2023fpx,Ferrante:2025ofe,Megias:2026qev,Brax:2019koq,Chaffey:2021tmj,Fichet:2022ixi,Fichet:2022xol,Fichet:2026nct}.
Ref.~\cite{Aoki:2023tjm} applied this continuum perspective to a simple
cosmological-collider toy model and found a universal damping of the
oscillatory signal in the squeezed limit.  Here we further develop this idea by
studying how a continuous spectrum modifies cosmological-collider signals and
by realizing it in an explicit higher-dimensional setup.  Closely related
ideas have also appeared in studies of strongly coupled, approximately
conformal, and unparticle-like sectors during inflation
\cite{Green:2013rd,Baumgart:2021ptt,Hubisz:2024xnj,Pimentel:2025rds,Yang:2025apy,Jiang:2025mlm}.

The basic physical mechanism is simple.  Different masses correspond to
different clock frequencies, and therefore the spectral average superposes
oscillations with different phases.  This dephases the sharp logarithmic
oscillation associated with an isolated particle and can smear it into a
smoother signal~\cite{Aoki:2023tjm}.  As we will see below, the behavior of
the spectral density near the threshold plays an important role in determining
the residual signal.

This mechanism is especially relevant when the four-dimensional spectrum is an
effective description of more microscopic degrees of freedom.  In
higher-dimensional theories, for example, four-dimensional particles arise as
Kaluza--Klein modes of bulk fields.  Depending on the geometry and boundary
conditions, the resulting spectrum need not consist only of isolated narrow
states; it may instead contain a continuum, or a very dense tower that
effectively behaves as a continuum. In this case, the absence of a sharp clock
signal does not necessarily imply the absence of heavy degrees of freedom
during inflation.  Rather, it may indicate that the relevant high-energy states
form a continuum whose different clock frequencies have been averaged over in
the primordial correlator.

We first develop a general description of this effect in a scalar toy model.
We introduce a massive scalar with a continuous spectral function and compute
the signal part of the inflaton four-point function mediated by its exchange.
The non-analytic part of the seed function can be written as an integral over
the mass parameter.  In the soft limit, this integral takes the form of a
Fourier--Laplace transform of the spectral density with respect to logarithmic
momentum variables.  This representation makes the smearing mechanism manifest.
For spectra with a nonzero principal-series threshold, an endpoint oscillation
can remain, although its amplitude is damped.  By contrast, if the continuum
starts at zero clock frequency, the endpoint oscillation itself disappears and
the signal becomes a smooth broad feature.

We then give a concrete realization of this idea in an inflationary version of
the Randall--Sundrum II (RS2) braneworld.  The original RS2 model provides a
brane-world realization of four-dimensional gravity in a non-compact warped
extra dimension~\cite{Randall:1999vf}.  Its de Sitter brane version, relevant
for inflation, contains a localized graviton zero mode and a continuum of
massive KK gravitons
\cite{Langlois:2000ns,Frolov:2002qm,Kobayashi:2005dd}. In this setup, the inflaton is localized
on a four-dimensional de Sitter brane embedded in a five-dimensional
anti-de Sitter bulk.  The tensor perturbations contain a localized massless
graviton zero mode and a continuum of KK gravitons separated by
a mass gap \(3H/2\), corresponding to the principal-series threshold
\(\mu=0\).  From the four-dimensional viewpoint, this continuum is a spectrum
of massive spin-2 particles in the principal series.  Importantly, the spectral
weight is not arbitrary: it is fixed by the five-dimensional wavefunctions and,
in particular, by their values on the brane.  We derive this KK spectral weight
and use it to construct the brane-to-brane propagator for the continuum KK
gravitons.

Finally, we study the contribution of the continuum KK gravitons to the
inflaton four-point function.  The coupling of each KK graviton to the
brane-localized inflaton is controlled by the value of its wavefunction at the
brane, and therefore the same quantity determines the effective spectral weight
appearing in the exchange diagram.  We show that the resulting seed function is
smooth in the soft region, without a persistent logarithmic oscillation.  This
provides an explicit higher-dimensional example in which the standard
single-particle cosmological-collider clock signal is smeared by a continuum of
massive spin-2 states.

The organization of this paper is as follows.  In Sec.~\ref{sec:Conti}, we
develop the general formalism for cosmological correlators with a continuous
mass spectrum and discuss the smearing of clock signals in a scalar toy model.
In Sec.~\ref{UV}, we introduce the RS2-like inflationary braneworld and derive
the continuum KK graviton spectrum.  We then compute the corresponding brane
spectral weight and apply it to the inflaton four-point function. In Sec.~\ref{sec:CFT_relation}, we discuss the relation to conserved conformal
stress-tensor continua. We conclude
in Sec.~\ref{sec:Conclusion}.  Technical details of the seed-function
calculation and the graviton equation of motion are provided in
Appendices~\ref{MBdetail} and~\ref{G_EOM}, respectively.

\paragraph{Conventions and notation.}
We use the mostly-plus metric signature in both four and five dimensions.
The four-dimensional exact de Sitter background is written as
\(\mathrm{d}s_4^2=-\mathrm{d}t^2+a^2(t)\mathrm{d}\mathbf{x}^2
=a^2(\tau)(-\mathrm{d}\tau^2+\mathrm{d}\mathbf{x}^2)\), with
\(a(t)=e^{Ht}\), \(a(\tau)=-1/(H\tau)\), and constant \(H\).
Bold symbols denote three-dimensional vectors.  Numerical subscripts on
momenta label external legs, while Latin indices \(i,j,\ldots\) denote spatial
components.  Thus, \(k_n\equiv|\mathbf{k}_n|\) and
\(k_{ni}\equiv(\mathbf{k}_n)_i\), where \(n\) labels an external leg.  We also
define \(k_{12}\equiv k_1+k_2\), \(k_{34}\equiv k_3+k_4\),
\(\mathbf{s}\equiv\mathbf{k}_1+\mathbf{k}_2\), and
\(s\equiv|\mathbf{s}|\).  A prime on a correlator indicates that the overall
factor \((2\pi)^3\delta^{(3)}(\sum_n\mathbf{k}_n)\) has been removed.
In the five-dimensional discussion, we use \(x^M=(x^\mu,y)\), or
\(x^M=(x^\mu,z)\) after introducing the conformal extra-dimensional
coordinate, where \(x^\mu=(t,\mathbf{x})\) or
\(x^\mu=(\tau,\mathbf{x})\).  Five-dimensional, four-dimensional, and spatial
indices are denoted by \(M,N,\ldots\), \(\mu,\nu,\ldots\), and
\(i,j,\ldots\), respectively.

%%%%%%%%%%%%%%%%%%%%%%%%%
\section{Cosmological Correlators with a Continuous Mass Spectrum}
\label{sec:Conti}

In this section, we develop a scalar toy model to describe the effects of a
continuous mass spectrum on cosmological correlators.  We denote the inflaton
fluctuation by \(\varphi\), and a massive scalar, possibly with a continuous
mass spectrum, by \(\sigma\).  For simplicity, we focus on scalar exchange; the
spectral part of the discussion generalizes to spinning particles, with
additional tensor structures.  An application to KK gravitons will be given in
Sec.~\ref{UV}.

%%%%%%%%%%%%%%%%%%%%%%%%%
\subsection{Mode Functions, Quantization, and Propagators}

Here we summarize the ingredients necessary for computing cosmological correlators with a continuous spectrum, such as mode functions and propagators.

\subsubsection*{Massless Scalar (Inflaton)}

A massless scalar field $\varphi$ is quantized as
\begin{align}
\varphi(\tau, \mathbf{x})
=
\int \frac{\mathrm{d}^3 \mathbf{k}}{(2 \pi)^3}
\left(
\varphi_k(\tau) a_{\mathbf{k}}
+
\varphi_k^*(\tau) a_{-\mathbf{k}}^{\dagger}
\right)
e^{i\mathbf{k} \cdot \mathbf{x}} .
\end{align}
Here, $a_{\mathbf{k}}$ and $a_{\mathbf{k}}^{\dagger}$ denote the annihilation and creation operators, respectively, satisfying the standard commutation relation
\begin{align}
[a_{\mathbf{k}}, a_{\mathbf{k}^{\prime}}^{\dagger}]
=
(2 \pi)^3
\delta^{(3)}\left(\mathbf{k}-\mathbf{k}^{\prime}\right).
\label{CR_1}
\end{align}
Assuming the Bunch--Davies (BD) vacuum, the mode function $\varphi_k(\tau)$ is given by
\begin{align}
\varphi_k(\tau)
=
\frac{H}{\sqrt{2 k^3}}
(1+i k \tau)
e^{-i k \tau}.
\end{align}

Since the inflaton always appears as an external line in the correlators considered in this work, it is convenient to define the bulk-to-boundary propagator
\begin{align}
K_{\mathrm{a}}(k, \tau)
=
\frac{H^2}{2 k^3}
(1-i\mathrm{a} k \tau)
e^{i \mathrm{a} k \tau},
\qquad
\mathrm{a}=\pm 1 .
\label{BB}
\end{align}

%%%%%%%%%%%%%%%%%%%%%%%%%%%%%%%%
\subsubsection*{Ordinary Massive Scalar}

Before turning to the main subject of this work, namely a massive scalar with a continuous mass spectrum, let us first review the case of a conventional massive scalar field in de Sitter spacetime. 
A massive scalar $\sigma$ with fixed mass $m$ is quantized as
\begin{align}
\sigma(\tau, \mathbf{x})
=
\int \frac{\mathrm{d}^3 \mathbf{k}}{(2 \pi)^3}
\left(
\sigma_{k}(\tau) b_{\mathbf{k}}
+
\sigma_{k}^*(\tau) b_{-\mathbf{k}}^{\dagger}
\right)
e^{i\mathbf{k} \cdot \mathbf{x}},
\label{q_s}
\end{align}
where the annihilation operator $b_{\mathbf{k}}$ and the creation operator $b_{\mathbf{k}}^\dagger$ satisfy commutation relations analogous to those in the massless case~\eqref{CR_1}. 
The mode function $\sigma_k(\tau)$ is given by
\begin{align}
\sigma_{k}(\tau)
=
-i
e^{i\left(\nu+\frac{1}{2}\right) \frac{\pi}{2}}
\frac{\sqrt{\pi}}{2}
H(-\tau)^{3 / 2}
H_{\nu}^{(1)}(-k \tau),
\label{mode_s}
\end{align}
where $H_{\nu}^{(1,2)}(z)$ denotes the Hankel function of the first or second kind, and the index $\nu$ is defined by
\begin{align}
\nu
\equiv
\sqrt{\frac{9}{4}-\left(\frac{m}{H}\right)^2}.
\end{align}

The parameter $\nu$ is real for $m<\frac{3}{2}H$, corresponding to the complementary series, while it becomes purely imaginary for $m>\frac{3}{2}H$, corresponding to the principal series. 
In the latter case, it is convenient to introduce a real parameter $\mu$ defined by
\begin{align}
\nu = i\mu,
\qquad
\mu \equiv \sqrt{\left(\frac{m}{H}\right)^2-\frac{9}{4}},
\end{align}
valid for $m>\frac{3}{2}H$.

The following relation is useful:
\begin{align}
\left(
e^{i \pi \nu / 2} H_\nu^{(1)}(z)
\right)^*
=
e^{-i \pi \nu / 2} H_\nu^{(2)}(z),
\label{formula_H}
\end{align}
which holds for real and purely imaginary $\nu$.

%%%%%%%%%%%%%%%%%%%%%%%%%
\subsubsection*{Massive Scalar with a Continuous Spectrum}

We now turn to a massive scalar field with a continuous mass spectrum in de Sitter spacetime. 
In the following, we closely follow Ref.~\cite{Aoki:2023tjm}, which generalized the flat-space results of Ref.~\cite{Csaki:2021gfm} to de Sitter spacetime. 
We will see that the toy model discussed in this section provides a useful description of the KK graviton continuum in the RS2 inflationary scenario.  

Given a spectral function $\rho(m^2)$, we define a massive scalar field with a continuous spectrum as
\begin{align}
\sigma(\tau, \mathbf{x})
=
\int \frac{\mathrm{d}m^2}{H^2} \sqrt{\rho\left(m^2\right)}
\int \frac{\mathrm{d}^3 \mathbf{k}}{(2 \pi)^3}
\left(
\sigma_{k,m}(\tau) b_{\mathbf{k},m}
+
\sigma_{k,m}^*(\tau) b_{-\mathbf{k},m}^{\dagger}
\right)
e^{i\mathbf{k} \cdot \mathbf{x}} .
\label{Q_sigma}
\end{align}
Here, the mode function is still given by Eq.~\eqref{mode_s}, but we attach the subscript $m$ to emphasize its mass dependence. 
The mass integral is taken over the range $m^2>0$. 
Since the mode function exhibits qualitatively different behavior depending on whether $m<\frac{3}{2}H$ or $m>\frac{3}{2}H$, it is convenient to decompose the integral as
\begin{align}
\int \mathrm{d} m^2
=
\int_0^{(3 H / 2)^2} \mathrm{d} m^2
+
\int_{(3 H / 2)^2}^{\infty} \mathrm{d} m^2,
\label{int_m}
\end{align}
where the first and second terms correspond to the complementary and principal series, respectively. 
The annihilation and creation operators now carry the mass index $m$ and satisfy
\begin{align}
\left[
b_{\mathbf{k},m},
b_{\mathbf{k}^{\prime},m^{\prime}}^{\dagger}
\right]
=
(2 \pi)^3
\delta^{(3)}\left(\mathbf{k}-\mathbf{k}^{\prime}\right)
H^2\delta(m^2-m^{\prime 2}).
\label{CR}
\end{align}

The spectral function $\rho(m^2)$ provides a new degree of freedom characterizing the theory. 
In particular, a delta-function spectral density reproduces the ordinary massive scalar field in Eq.~\eqref{q_s}. 
More general choices lead to various novel effects, which will be discussed in detail in the subsequent sections. 
To simplify the notation, we introduce the abbreviation
\begin{align}
\int_{m^2}
\equiv
\int \frac{\mathrm{d} m^2}{H^2} \rho(m^2).
\label{SI}
\end{align}

Finally, we introduce the bulk-to-bulk propagators for $\sigma$, denoted by $D_{\mathrm{a}\mathrm{b}}$, where $\mathrm{a},\mathrm{b}=\pm$ are Schwinger--Keldysh indices; see Ref.~\cite{Chen:2017ryl} for details. 
Explicitly, they are given by
\begin{align}
D_{-+}\left(k ; \tau_1, \tau_2\right)
&=
\int_{m^2}
\sigma_{k,m}\left(\tau_1\right)
\sigma_{k,m}^*\left(\tau_2\right)
\nonumber\\
&=
\int_{m^2}
\frac{H^2 \pi}{4}
\left(-\tau_1\right)^{3 / 2}
\left(-\tau_2\right)^{3 / 2}
H_{\nu}^{(1)}\left(-k \tau_1\right)
H_{\nu}^{(2)}\left(-k \tau_2\right),
\label{SK_-+}
\\
D_{+-}\left(k ; \tau_1, \tau_2\right)
&=
D_{-+}\left(k ; \tau_2, \tau_1\right)
=
\left(
D_{-+}\left(k ; \tau_1, \tau_2\right)
\right)^*,
\\
D_{\pm \pm}\left(k ; \tau_1, \tau_2\right)
&=
D_{\mp \pm}\left(k ; \tau_1, \tau_2\right)
\theta\left(\tau_1-\tau_2\right)
+
D_{\pm \mp}\left(k ; \tau_1, \tau_2\right)
\theta\left(\tau_2-\tau_1\right),
\label{SK_++}
\end{align}
where we used Eq.~\eqref{formula_H}.

%%%%%%%%%%%%%%%%%%%%%%%%%%%%%%%
\subsection{Scalar Exchange with a Continuous Spectrum}

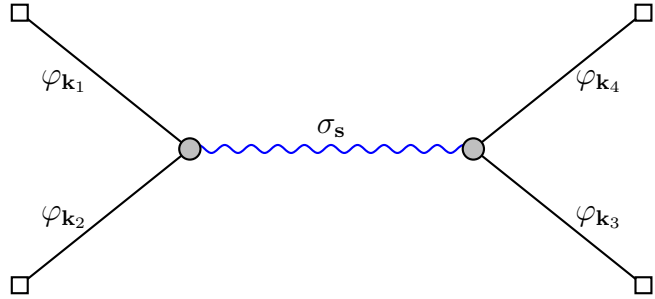
\begin{figure}[H]
    \centering

\begin{tikzpicture}[thick, scale=1.5]

%--- style definitions ---
\tikzset{
    ext/.style={draw, rectangle, minimum size=6pt, inner sep=0pt},
    vertex/.style={
        draw,
        circle,
        minimum size=8pt,
        inner sep=0pt,
        fill=lightgray
    }
}

%--- vertices ---
\coordinate (v1) at (0,0);
\coordinate (v2) at (2.5,0);

%--- external endpoints ---
\coordinate (k1) at (-1.5,1.2);
\coordinate (k2) at (-1.5,-1.2);
\coordinate (k4) at (4,1.2);
\coordinate (k3) at (4,-1.2);

%--- external lines ---
\draw (k1) -- (v1);
\draw (k2) -- (v1);
\draw (v2) -- (k4);
\draw (v2) -- (k3);

%--- endpoint squares ---
\node[ext,fill=white] at (k1) {};
\node[ext,fill=white] at (k2) {};
\node[ext,fill=white] at (k3) {};
\node[ext,fill=white] at (k4) {};

%--- internal line (blue) ---
\draw[
    blue,
    thick,
    decorate,
    decoration={snake, amplitude=1.5pt, segment length=10pt}
] (v1) -- (v2);

%--- vertices ---
\node[vertex] at (v1) {};
\node[vertex] at (v2) {};

%--- labels ---
\node[left] at (-0.8,0.6) {$\varphi_{\mathbf{k}_1}$};
\node[left] at (-0.8,-0.6) {$\varphi_{\mathbf{k}_2}$};

\node[right] at (3.3,0.6) {$\varphi_{\mathbf{k}_4}$};
\node[right] at (3.3,-0.6) {$\varphi_{\mathbf{k}_3}$};

\node[above] at (1.25,0) {$\sigma_{\mathbf{s}}$};

\end{tikzpicture}

\caption{$s$-channel diagram for the inflaton four-point correlator. The blue wavy line denotes the massive scalar field with a continuous mass spectrum.}
\label{Fig:SK_diagrams}
\end{figure}

Having introduced the basic ingredients, we now specify the target observable.
We mainly focus on the four-point function of the inflaton fluctuation
\(\varphi\) mediated by the exchange of~\(\sigma\).  The \(s\)-channel diagram
is shown in Fig.~\ref{Fig:SK_diagrams}, while the full correlator also contains
the momentum-permuted \(t\)- and \(u\)-channel contributions.

The diagram in Fig.~\ref{Fig:SK_diagrams} can, for example, arise from the following cubic interaction:
\begin{align}
S_{\rm cubic}
=
\int \mathrm{d} \tau \, \mathrm{d}^3 \mathbf{x}
\;
\frac{a(\tau)^2}{2}
\sigma \varphi^{\prime 2},
\label{3vertex}
\end{align}
where the prime denotes differentiation with respect to the conformal time $\tau$. 
This interaction is meant as an illustrative example.  In the expression below,
we will also introduce a generalized seed function in which the powers of
\(-\tau_i\) at the two vertices are left arbitrary (see Eq.~\eqref{def_seed}).  This notation allows us to
treat other interaction-types and time dependences in a unified way; see,
e.g., Ref.~\cite{Qin:2023ejc} for more discussions.

Applying the Schwinger--Keldysh (SK) diagrammatic rules~\cite{Chen:2017ryl} to the four-point correlator generated by the interaction~\eqref{3vertex}, evaluated at the future boundary $\tau=0$, we obtain
\begin{align}
\nonumber
\left\langle
\varphi_{\mathbf{k}_1}
\varphi_{\mathbf{k}_2}
\varphi_{\mathbf{k}_3}
\varphi_{\mathbf{k}_4}(0)
\right\rangle^{\prime}
=
&\;
\int_{-\infty}^0 \mathrm{d} \tau_1 \int_{-\infty}^0\mathrm{d} \tau_2
\left(-H\tau_1\right)^{-2}
\left(-H\tau_2\right)^{-2}
\sum_{\mathrm{a}, \mathrm{b}= \pm}
(-\mathrm{a}\mathrm{b})
\\
\nonumber &\times
K_{\mathrm{a}}^{\prime}\left(k_1 ; \tau_1\right)
K_{\mathrm{a}}^{\prime}\left(k_2 ; \tau_1\right)
K_{\mathrm{b}}^{\prime}\left(k_3 ; \tau_2\right)
K_{\mathrm{b}}^{\prime}\left(k_4 ; \tau_2\right)
D_{\mathrm{a}\mathrm{b}}\left(s ; \tau_1, \tau_2\right)\\
&+t+u,
\label{4pt_1}
\end{align}
where the prime on the left-hand side indicates that the overall momentum-conservation factor
$(2\pi)^3\delta^{(3)}(\mathbf{k}_1+\mathbf{k}_2+\mathbf{k}_3+\mathbf{k}_4)$
has been omitted. 
The bulk-to-boundary propagator $K_{\mathrm{a}}$ and the bulk-to-bulk propagator $D_{\mathrm{a}\mathrm{b}}$ are defined in Eq.~\eqref{BB} and Eqs.~\eqref{SK_-+}--\eqref{SK_++}, respectively. 
Here and in the following, we define
\begin{align}
\mathbf{s}\equiv\mathbf{k}_1+\mathbf{k}_2,
\qquad
s\equiv |\mathbf{s}|,
\qquad
k_{12}\equiv k_1+k_2, \qquad
k_{34}\equiv k_3+k_4  .
\end{align}
The $t$- and $u$-channel contributions are implicit.

Furthermore, by substituting the explicit expression for $K_{\mathrm{a}}$, Eq.~\eqref{4pt_1} can be rewritten as
\begin{align}
\left\langle
\varphi_{\mathbf{k}_1}
\varphi_{\mathbf{k}_2}
\varphi_{\mathbf{k}_3}
\varphi_{\mathbf{k}_4}(0)
\right\rangle^{\prime}
=&\;
\frac{H^4}{16 k_1 k_2 k_3 k_4}
\sum_{\mathrm{a}, \mathrm{b}= \pm}
(-\mathrm{a}\mathrm{b})
\int_{-\infty}^0
\mathrm{d} \tau_1 \int_{-\infty}^0\mathrm{d} \tau_2
\left(-\tau_1\right)^{-2}
\left(-\tau_2\right)^{-2}
\nonumber\\
&\times
e^{i\mathrm{a} k_{12} \tau_1+i\mathrm{b} k_{34} \tau_2}
D_{\mathrm{a}\mathrm{b}}\left(s ; \tau_1, \tau_2\right)
+t+u
\nonumber\\
=&\;
\frac{H^6}{16 k_1 k_2 k_3 k_4s^5}
\sum_{\mathrm{a}, \mathrm{b}= \pm}
\mathcal{I}_{\mathrm{a}\mathrm{b}}^{00}
+t+u,\label{4pt}
\end{align}
where we have introduced the seed function
\begin{align}
\mathcal{I}_{\mathrm{a}\mathrm{b}}^{p_1 p_2}
\equiv
\frac{s^{5+p_1+p_2}}{H^2}
(-\mathrm{a}\mathrm{b})
\int_{-\infty}^0
\mathrm{d} \tau_1 \int_{-\infty}^0\mathrm{d} \tau_2
\left(-\tau_1\right)^{p_1}
\left(-\tau_2\right)^{p_2}
e^{i\mathrm{a} k_{12} \tau_1+i\mathrm{b} k_{34} \tau_2}
D_{\mathrm{a}\mathrm{b}}\left(s ; \tau_1, \tau_2\right).
\label{def_seed}
\end{align}
Here, \(p_i\) \((i=1,2)\) parametrize the powers of \(-\tau_i\) associated with
the two interaction vertices.  In the example of Eq.~\eqref{3vertex}, one has
\(p_1=p_2=0\).  More general interactions can be incorporated by choosing
different values of \(p_i\).  The problem is therefore reduced to evaluating the
seed function~\eqref{def_seed}.
Lower-point correlators, such as two- and three-point correlators, can be obtained by taking one or more external momenta to zero, although this limit is nontrivial because of apparent cancellations among divergent terms~\cite{Qin:2023ejc,Aoki:2023wdc}.

It turns out that the momentum dependence of the seed function enters only through the following two combinations:
\begin{align}
r_1 \equiv \frac{s}{k_{12}},
\qquad
r_2 \equiv \frac{s}{k_{34}}.
\end{align}
The evaluation of the seed function for the standard case of a discrete mass spectrum has been discussed extensively in the literature. We follow the Mellin--Barnes approach
to the Hankel functions~\cite{Qin:2022lva,Qin:2022fbv}, which allows us to
perform the time integrals while keeping the mass integral explicit.  The
technical details are summarized in Appendix~\ref{MBdetail}.
The subsequent calculation then proceeds in close analogy with the ordinary discrete-mass case.

As in the conventional case, the seed function can be decomposed into a non-analytic part, which contains the physical signal and is oscillatory in the principal series, and a featureless analytic part, which constitutes the background:
\begin{align}
\mathcal{I}_{\rm Total}
\equiv
\sum_{\mathrm{a}, \mathrm{b}= \pm}
\mathcal{I}_{\mathrm{a}\mathrm{b}}^{p_1 p_2}
=
\mathcal{I}_{\rm Signal}
+
\mathcal{I}_{\rm BG}.\label{Seed_dec}
\end{align}

We leave the detailed derivation to Appendix~\ref{MBdetail}, and here present the result valid for $r_1<r_2$:\footnote{The result for $r_1>r_2$ can be obtained by the replacements $r_1\leftrightarrow r_2$ and $p_1\leftrightarrow p_2$.}
\begin{align}
\mathcal{I}_{\rm Signal}
=
\int_{m^2}
\mathcal{C}_{\nu}^{p_1 p_2}
\left[
\left(\frac{r_1 r_2}{4}\right)^{\nu}
\mathbf{F}_{\nu}^{p_1}\left(r_1\right)
\mathbf{F}_{\nu}^{p_2}\left(r_2\right)
+
\left(\frac{r_1}{r_2}\right)^\nu
\mathbf{F}_\nu^{p_1}\left(r_1\right)
\mathbf{F}_{-\nu}^{p_2}\left(r_2\right)
+
(\nu\rightarrow -\nu)\right]
,
\label{I_signal}
\end{align}
where
\begin{align}
\mathcal{C}_{\nu}^{p_1 p_2}
&\equiv
\frac{1}{2 \pi}
\left[
\cos \left(\frac{\pi}{2}(p_1-p_2)\right)
+
\sin \left(\frac{\pi}{2}\left(p_1+p_2+2 \nu\right)\right)
\right],
\\
\mathbf{F}_{\nu}^p(r)
&\equiv
r^{5 / 2+p}
\Gamma\left(\frac{5}{2}+p+ \nu\right)
\Gamma\left(- \nu\right)
{}_2 \mathrm{F}_1
\left[
\left.
\begin{array}{c}
\frac{5}{4}+\frac{p}{2}+\frac{\nu}{2},
\frac{7}{4}+\frac{p}{2}+\frac{\nu}{2}
\\
1+ \nu
\end{array}
\right\rvert\,
r^2
\right].
\label{def_F^p_nu}
\end{align}
Here, ${}_2 \mathrm{F}_1$ denotes the hypergeometric function, and $\nu$ should be understood as either real or purely imaginary, depending on the integration region of the mass spectrum. 
The terms proportional to $(r_1r_2/4)^{\nu}$ and $(r_1/r_2)^{\nu}$ in Eq.~\eqref{I_signal} correspond to the local and nonlocal signals, respectively~\cite{Tong:2021wai}. 
The background contribution has a more complicated structure, involving double infinite sums; its explicit expression is given in Eq.~\eqref{I_BG}.

In the hierarchical soft limit $r_1\ll r_{2}\ll 1$, 
the signal part typically dominates over the background contribution. 
Moreover, the hypergeometric function in Eq.~\eqref{def_F^p_nu} 
can be approximated by unity in that case. 
In the following subsection, we focus on this signal part in the soft limit, 
or, more precisely, on its normalized version,
\begin{align}
\mathcal{F}_{\rm Signal}
\equiv
\frac{
\mathcal{I}_{\rm Signal}
}{
-4\pi r_1^{\frac{5}{2}+p_1}r_2^{\frac{5}{2}+p_2}
}.
\label{def_NF}
\end{align}

%%%%%%%%%%%%
\subsection{Principal-Series Continua and Threshold Behavior}

Of particular interest is the case in which the continuous spectrum turns on above a nonzero threshold scale, corresponding to a gapped continuum. In this subsection, we consider the case in which the gap scale lies above $3H/2$, so that all massive states belong to the principal series:
\begin{align}
\rho(m^2)
=
\varrho(m^2)\theta(m^2-m_0^2),
\qquad
m_0>\frac{3H}{2},
\end{align}
where $m_0$ denotes the gap scale and $\theta(z)$ is the Heaviside step function. 

The spectral integral~\eqref{SI} can then be rewritten as
\begin{align}
\int_{m^2}
&=
\int_{0}^{\infty}
\frac{\mathrm{d} m^2}{H^2}
\varrho(m^2)
\theta(m^2-m_0^2)
=
\int_{m_0^2}^{\infty}
\frac{\mathrm{d} m^2}{H^2}
\varrho(m^2)
\nonumber\\
&=
\int_{\mu_0}^{\infty}
\mathrm{d} \mu\,
2 \mu\,
\varrho(\mu),
\end{align}
where
\begin{align}
\mu_0
\equiv
\sqrt{
\left(\frac{m_0}{H}\right)^2
-\frac{9}{4}
}
>0.
\end{align}
In the last equality, we used $m^2/H^2=\mu^2+9/4$ and, by a slight abuse of notation, defined
\begin{align}
\varrho(\mu)
\equiv
\varrho\left(H^2\left(\mu^2+\frac{9}{4}\right)\right).
\end{align}
%The limiting case \(\mu_0=0\), which is relevant for the RS2-like setup discussed in Sec.~\ref{UV}, will be obtained by taking the threshold to the principal-series edge.

Performing the spectral integral in Eq.~\eqref{I_signal} analytically is in general nontrivial. 
However, for states with $\mu\gtrsim \mathcal{O}(1)$, the Stirling approximation
\begin{align}
\Gamma(a \pm i b)
\simeq
\sqrt{2 \pi}\,
b^{a-1 / 2}
e^{-\pi b / 2}
e^{\pm i(b \log b+\pi a / 2)},
\qquad
a,b \in \mathbb{R},
\qquad
b \gg 1 ,
\end{align}
already provides a useful approximation. 
One can therefore approximate the integrand of Eq.~\eqref{def_NF} as
\begin{align}
\mathcal{F}_{\rm Signal}
&\sim
\int_{\mu_0}^{\infty}
\mathrm{d} \mu\,
\varrho(\mu)
\mu^{4+p_1+p_2}
e^{-\pi \mu}
\left[
\cos \left(\mu L_1\right)
+
\sin \left(\mu L_2-p_2 \pi\right)
\right]
\nonumber\\
&=
\operatorname{Re}\left[\Phi(L_1)\right]
+
\operatorname{Im}\left[\Phi(L_2)e^{-ip_2\pi}\right],
\label{F_signal}
\end{align}
where we have defined
\begin{align}
\Phi(L)
\equiv
\int_{\mu_0}^{\infty}
\mathrm{d} \mu\,
\varrho(\mu)
\mu^{4+p_1+p_2}
e^{-\pi \mu}
e^{i\mu L},
\label{def_F}
\end{align}
with
\begin{align}
L_1\equiv \log \frac{r_1 r_2}{4},
\qquad
L_2\equiv \log \frac{r_1}{r_2}.
\end{align}

We note the expression~\eqref{F_signal} should be understood as a soft-limit, large-$\mu$ approximation to the signal part. 
It captures the logarithmic oscillations and their envelope, which are the main features of interest in the following discussion. In the following, we investigate several examples with different choices of the spectral function $\varrho(\mu)$.

%%%%%%%%%%%%%%%%%%%%%%%%%%%%%%%%%%%%%%
\subsubsection*{Example: Delta function}

As a warm-up example, let us consider a delta-function spectral density:
\begin{align}
\varrho(\mu)
=
\delta(\mu-\mu_*),
\qquad
\mu_*>\mu_0 .
\end{align}
Then Eq.~\eqref{F_signal} reduces, up to an overall normalization, to the
standard fixed-mass cosmological-collider signal with oscillation frequency
\(\mu_*\).  Since the spectral weight is localized at a single frequency, there
is no dephasing or smearing effect.

%%%%%%%%%%%%%%%%%%%%%%%%%%%%
\subsubsection*{Example: Constant/power-law continuum}

The next example is a power-law spectral density,
\begin{align}
\varrho(\mu)=\mu^{\alpha},
\qquad
\alpha\in \mathbb{R},
\label{rho_a}
\end{align}
where $\alpha=0$ corresponds to a constant spectral function. 
This form provides a simple parametrization of continuum spectral densities.

In this case, we can explicitly perform the spectral integral in Eq.~\eqref{def_F} and obtain
\begin{align}
\Phi(L)
&=
\int_{\mu_0}^{\infty}
\mathrm{d} \mu\,
\mu^{4+p_1+p_2+\alpha}
e^{-\pi \mu}
e^{i\mu L}
\nonumber\\
&=
\frac{
\Gamma\left[5+p_1+p_2+\alpha,\mu_0(\pi-i L)\right]
}{
(\pi-i L)^{5+p_1+p_2+\alpha}
},
\label{F_PL}
\end{align}
where $\Gamma[a,z]$ denotes the incomplete Gamma function.

Moreover, for $\mu_0 \gtrsim 1$, we may use
\begin{align}
\Gamma[a, z]
\simeq
z^{a-1} e^{-z},
\qquad
|z|\gtrsim 1,
\end{align}
and the signal~\eqref{F_signal} can then be approximated as
\begin{align}
\mathcal{F}_{\rm Signal}
\sim
\mathcal{A}
\left[
\frac{
\cos \left(\mu_0L_1+\operatorname{arctan}\frac{L_1}{\pi}\right)
}{
\sqrt{\pi^2+L_1^2}
}
+
\frac{
\sin \left(\mu_0L_2+\operatorname{arctan}\frac{L_2}{\pi}-p_2\pi\right)
}{
\sqrt{\pi^2+L_2^2}
}
\right],
\label{Signal_alpha}
\end{align}
where
\begin{align}
\mathcal{A}
\equiv
\mu_0^{4+p_1+p_2+\alpha}e^{-\pi \mu_0}.
\end{align}

Thus we find that the oscillation frequency is fixed by the end point frequency $\mu_0$. Furthermore, the signal exhibits a scaling suppression proportional to $|L|^{-1}$ for $|L|\gg 1$, which corresponds to the soft limit. 
Remarkably, this suppression depends neither on $\alpha$ in Eq.~\eqref{rho_a} nor on the interaction type specified by $p_{1,2}$. 
%This universal damping behavior was already pointed out in Ref.~\cite{Aoki:2023tjm}; here, we explicitly demonstrate it using the analytic expression above.

\begin{figure}[H]
 \begin{minipage}{0.5\hsize}
  \begin{center}
   \includegraphics[width=80mm]{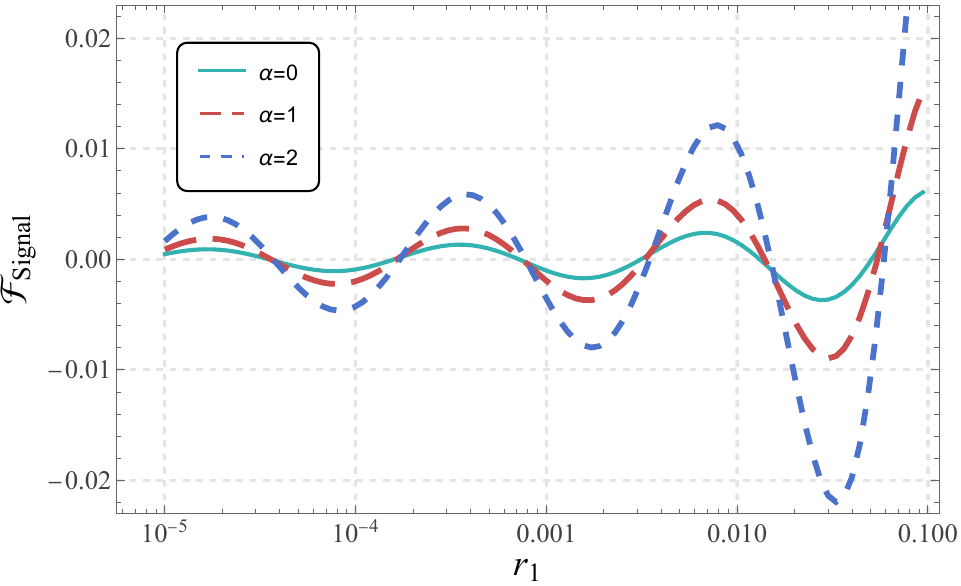}
  \end{center}
 \end{minipage}
 \begin{minipage}{0.5\hsize}
  \begin{center}
   \includegraphics[width=80mm]{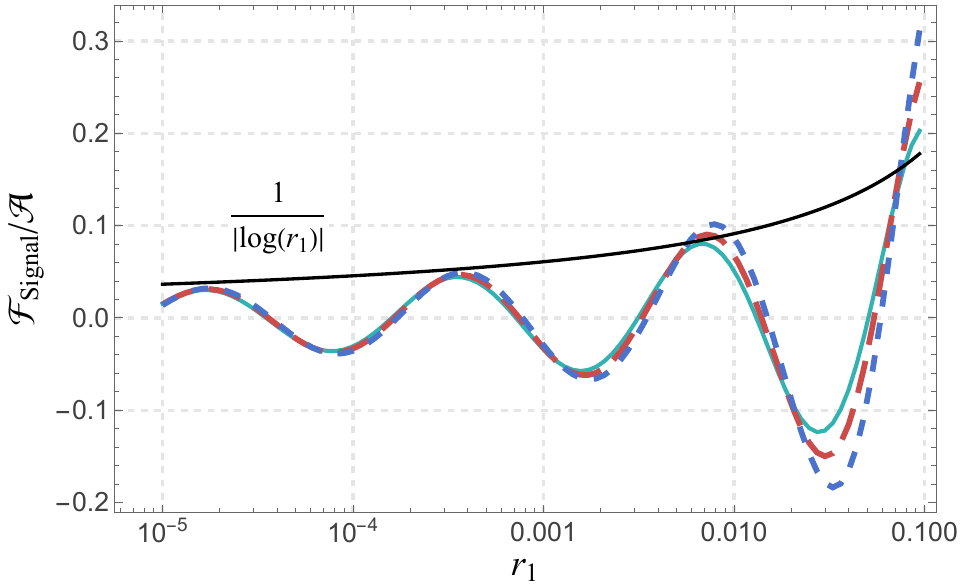}
  \end{center}
 \end{minipage}
\caption{
Signal part of the normalized seed~\eqref{def_NF} for the power-law spectral
function with \(\mu_0=2\) and \((p_1,p_2)=(0,0)\).
\textit{Left panel}: Dependence on the momentum ratio \(r_1\) for fixed
\(r_2=2/3\) and several choices of \(\alpha=\{0,1,2\}\).
\textit{Right panel}: The same quantity divided by the mass-dependent prefactor
\(\mathcal A=\mu_0^{4+p_1+p_2+\alpha}e^{-\pi\mu_0}\).  The black curve is
proportional to \(1/|\log r_1|\) and is shown as a guide to the eye.
}
\label{FigPL}
\end{figure}

In the left panel of Fig.~\ref{FigPL}, we show the signal part of the normalized seed~\eqref{def_NF} for several values of $\alpha=0,1,2$, plotted as a function of the momentum ratio $r_1$ with fixed $r_2$. 
Small $r_1$ corresponds to the squeezed configuration. 
One can clearly observe the damping of the oscillatory signal in the squeezed limit, in agreement with the analytic estimate discussed above. 
The right panel shows the same quantity further rescaled by the mass-dependent prefactor $\mathcal{A}$. 
This makes the universal $1/|\log r_1|$ behavior manifest, independently of the value of $\alpha$.

%%%%%%%%%%%%%%%%%%%%%%%%%%%%%%%%%%%%%%%%%%%%%%%%%%%    
\subsubsection*{Example: Threshold power-law}

The next example is the case in which the spectral density turns on smoothly from the threshold:
\begin{align}
\varrho(\mu)
=
\left(\mu-\mu_0\right)^\alpha,
\qquad
\alpha\in \mathbb{R}.
\end{align}
For the threshold behavior to be integrable, one should take $\alpha>-1$. 
In this case, the spectral integral in Eq.~\eqref{def_F} can be evaluated as
\begin{align}
\Phi(L)
&=
\int_{\mu_0}^{\infty}
\mathrm{d} \mu\,
\mu^{4+p_1+p_2}
(\mu-\mu_0)^{\alpha}
e^{-\pi \mu}
e^{i\mu L}
\nonumber\\
&=
\mu_0^{5+p_1+p_2+\alpha}
e^{-(\pi -i L) \mu_0}
\Gamma (\alpha +1)
U(\alpha +1, 6+p_1+p_2+\alpha,(\pi -i L)\mu_0),
\label{F_TPL}
\end{align}
where $U(a,b,z)$ denotes the Tricomi confluent hypergeometric function. 
Note that Eq.~\eqref{F_TPL} agrees with the power-law result in Eq.~\eqref{F_PL} in the limit $\mu_0=0$ and/or $\alpha=0$, as expected.\footnote{This formal limit should not be interpreted as a controlled approximation to
the exact seed at an endpoint \(\mu_0=0\), since the fixed-mass kernel used in
Eq.~\eqref{def_F} was obtained using the large-\(\mu\) approximation.}

For $\mu_0\gtrsim 1$, however, the behavior differs from that of the power-law spectral function. 
Using
\begin{align}
U(a,b,z)\simeq z^{-a},
\qquad
|z|\gtrsim 1,
\end{align}
the signal~\eqref{F_signal} can be approximately written as
\begin{align}
\mathcal{F}_{\rm Signal}
\sim
\mathcal{A}
\left[
\frac{
\cos \left(\mu_0L_1+(\alpha+1)\operatorname{arctan}\frac{L_1}{\pi}\right)
}{
(\pi^2+L_1^2)^{\frac{\alpha+1}{2}}
}
+
\frac{
\sin \left(\mu_0L_2+(\alpha+1)\operatorname{arctan}\frac{L_2}{\pi}-p_2\pi\right)
}{
(\pi^2+L_2^2)^{\frac{\alpha+1}{2}}
}
\right],
\label{Signal_alpha_TPL}
\end{align}
where now
\begin{align}
\mathcal{A}
\equiv
\mu_0^{4+p_1+p_2}
\Gamma(\alpha+1)
e^{-\pi \mu_0}.
\end{align}

Compared with the power-law result in Eq.~\eqref{Signal_alpha}, we find an $\alpha$-dependent suppression proportional to $|L|^{-(\alpha+1)}$ in the squeezed configuration. 
This additional suppression originates from the smooth threshold behavior of the spectral density near $\mu=\mu_0$ (see the discussion around Eq.~\eqref{eq:endpoint_smearing}).

\begin{figure}[H]
  \begin{center}
   \includegraphics[width=100mm]{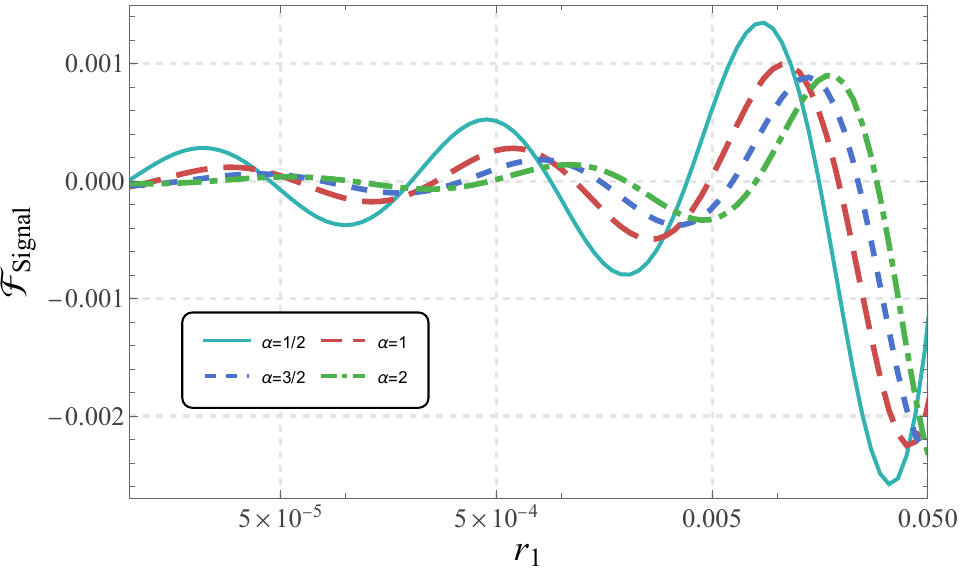}
  \end{center}
\caption{
Signal part of the normalized seed~\eqref{def_NF} for the threshold power-law spectral function with $\mu_0=2$ and $(p_1,p_2)=(0,0)$, shown as a function of $r_1$ for fixed $r_2=2/3$. 
Different curves correspond to $\alpha=1/2,1,3/2,2$. 
The envelope becomes more strongly damped for larger $\alpha$, in agreement with the scaling $|\log r_1|^{-(\alpha+1)}$.
}
\label{FigTPL}
\end{figure}

Figure~\ref{FigTPL} shows the signal part of the normalized seed~\eqref{def_NF} for several values of $\alpha=1/2,1,3/2,2$, plotted as a function of the momentum ratio $r_1$ with fixed $r_2$.

%%%%%%%%%%%%%%%%%%%%%%%%%%%%%%%%%%%%%%%%%%%%%%%
%%%%%%%%%%%%%%%%%%%%%%%%%%%%%%%%%%%%%%%%%%%%%%%
\subsubsection*{Example: Narrow Gaussian}

Finally, we consider a Gaussian spectral function,
\begin{align}
\varrho(\mu)
=
\exp\left[
-\frac{(\mu-\mu_*)^2}{\Delta^2}
\right],
\qquad
\mu_*\in \mathbb{R},
\qquad
\Delta>0 .
\label{rho_NG}
\end{align}
Here $\mu_*$ denotes the center of the Gaussian and $\Delta$ its width.
The corresponding function $\Phi(L)$ is
\begin{align}
\Phi(L)
=
\int_{\mu_0}^{\infty}
\mathrm{d}\mu\,
\mu^{4+p_1+p_2}
\exp\left[
-\frac{(\mu-\mu_*)^2}{\Delta^2}
\right]
e^{-(\pi-iL)\mu}.
\label{Phi_NG_def}
\end{align}

The behavior of the signal depends on the position of the Gaussian peak
relative to the threshold.  When the peak lies well above the threshold,
\(\mu_*-\mu_0\gtrsim \Delta\), the integral is controlled by the finite-width
distribution around \(\mu_*\).  In the regime where the lower endpoint is
negligible, one may extend the lower limit to \(-\infty\).  This gives,
schematically,
\begin{align}
\Phi(L)
\sim
e^{-\pi\mu_*}
e^{i\mu_*L}
e^{-\Delta^2L^2/4}
\times
\text{polynomial in }L .
\label{Phi_NG_saddle}
\end{align}
Here we have omitted an overall normalization and a mild phase shift that do
not affect the Gaussian damping.  Thus the signal resembles a single-particle
clock with frequency \(\mu_*\), but its amplitude is damped by the dephasing of
nearby frequencies in the Gaussian distribution.

On the other hand, when the Gaussian peak lies below the threshold,
$\mu_*<\mu_0$, the saddle is outside the integration domain.  In this case the
integral is controlled by the lower endpoint $\mu=\mu_0$.  Writing
\begin{align}
\delta_0\equiv \mu_0-\mu_* >0 ,
\end{align}
and expanding the integrand around $\mu=\mu_0$, the leading endpoint estimate gives
\begin{align}
\Phi(L)
\simeq
\frac{
\mu_0^{4+p_1+p_2}
e^{-\delta_0^2/\Delta^2}
e^{-\mu_0(\pi-iL)}
}{
\pi-iL+\frac{2\delta_0}{\Delta^2}
}.
\label{Phi_NG_endpoint}
\end{align}
The Gaussian suppression factor $e^{-\Delta^2L^2/4}$ is absent in this
endpoint-dominated regime.  Instead, the signal is an endpoint ringing with
frequency $\mu_0$, whose amplitude is suppressed by the Gaussian tail evaluated
at the threshold.

\begin{figure}[H]
  \begin{center}
   \includegraphics[width=100mm]{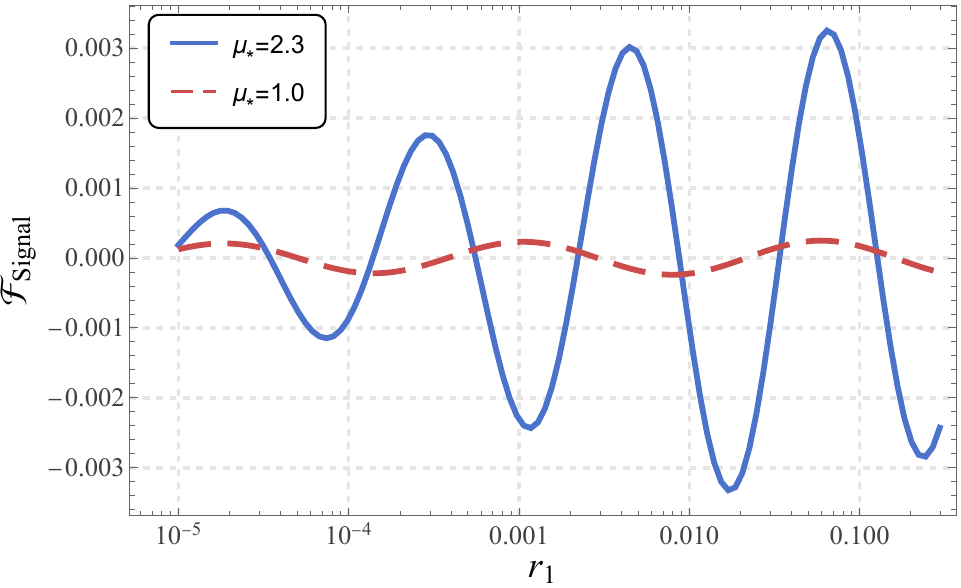}
  \end{center}
\caption{
Signal part of the normalized seed~\eqref{def_NF} for the Gaussian spectral
function~\eqref{rho_NG} with $\mu_0=1.5$, $\Delta=0.3$, and
$(p_1,p_2)=(0,0)$, shown as a function of $r_1$ for fixed $r_2=2/3$.
The blue curve corresponds to $\mu_*=2.3$, for which the Gaussian peak lies above
the threshold and gives a visible oscillatory signal.  The red dashed curve corresponds to \(\mu_*=1.0\), for which the peak lies below
the threshold.  In this case, the signal is dominated by the Gaussian tail near
the lower endpoint \(\mu=\mu_0\).  For the parameters shown here,
\((\mu_0-\mu_*)/\Delta\simeq1.67\), and the tail contribution is strongly
suppressed.
}
\label{FigNG}
\end{figure}

Figure~\ref{FigNG} illustrates these two regimes.  The blue curve corresponds to a
Gaussian peak above the threshold, for which a visible oscillatory signal remains.
The red dashed curve corresponds to a peak below the threshold.  In
this case, the signal is dominated by the Gaussian tail near the lower endpoint
\(\mu=\mu_0\), and its amplitude is suppressed by the factor
\(\exp[-(\mu_0-\mu_*)^2/\Delta^2]\).  For the parameter choice shown in the
figure, this suppression is sizable.

%%%%%%%%%%%%%%%%%%%%%%%%%%%%%%%%%%%%%%%%%%%%%%%%%%%%%%%%%%%%%%%%%%%%%%%%%%
\subsubsection*{Endpoint Behavior and Continuum Smearing}

Within the soft-limit and large-\(\mu\) approximation used above, the examples
can be understood in a unified way from the representation
\begin{align}
  \Phi(L)
  =
  \int_{\mu_0}^{\infty} \mathrm{d}\mu\,
  W(\mu)\,e^{i\mu L},
  \qquad
  W(\mu)
  \equiv
  \varrho(\mu)\,\mu^{4+p_1+p_2}e^{-\pi\mu}.
  \label{eq:Phi_as_smearing}
\end{align}
For a single massive particle, the spectral weight is localized at one value of
\(\mu\), and the signal contains a definite clock frequency.  By contrast, a
continuum spectrum gives a coherent superposition of clock signals with
different frequencies.  In the logarithmic momentum variable \(L\), these
different frequency components dephase as \(|L|\) increases.  This dephasing is
what we refer to as continuum smearing.

The large-\(|L|\) behavior is controlled by the structure of the effective
spectral weight near the lower endpoint $\mu_0$.  Suppose that near the threshold
\begin{align}
  \varrho(\mu)\,\mu^{4+p_1+p_2}
  =
  c_\beta\,(\mu-\mu_0)^\beta+\cdots,
  \qquad
  \beta>-1 .
\end{align}
Then the endpoint contribution behaves as
\begin{align}
  \Phi(L)
  \sim
  c_\beta\,
  e^{-\pi\mu_0}
  e^{i\mu_0 L}
  \frac{\Gamma(\beta+1)}{(\pi-iL)^{\beta+1}},
  \qquad
  |L|\gg 1 .
  \label{eq:endpoint_smearing}
\end{align}
This formula explains the examples considered above.  A power-law spectrum
\(\varrho(\mu)=\mu^\alpha\) with a nonzero threshold \(\mu_0>0\) has a sharp
edge, corresponding to \(\beta=0\), and therefore gives a universal
\(1/|L|\) damping while retaining an endpoint oscillation with frequency
\(\mu_0\).  A threshold power-law spectrum,
\(\varrho(\mu)\propto(\mu-\mu_0)^\alpha\), has \(\beta=\alpha\) for
\(\mu_0>0\) and hence exhibits the stronger damping
\(1/|L|^{\alpha+1}\).

The limiting case \(\mu_0=0\) is qualitatively different.  Since the endpoint
frequency vanishes, \(e^{i\mu_0L}=1\), no persistent logarithmic oscillation
with a nonzero endpoint frequency remains.  The precise large-\(|L|\) damping
is controlled by the small-\(\mu\) behavior of the full effective kernel
constructed from the exact fixed-mass seed, and cannot in general be inferred
from the large-\(\mu\) approximation used above. This case is relevant for the RS2-like KK continuum discussed in
Sec.~\ref{UV}, where the smooth non-oscillatory behavior is evaluated using
the exact fixed-mass signal and the KK spectral weight.

Thus continuum smearing does not necessarily imply the complete disappearance
of oscillations.  A nonzero endpoint \(\mu_0>0\) can leave a damped endpoint
clock, whereas a continuum starting at \(\mu_0=0\) has no nonzero endpoint
frequency.

%%%%%%%%%%%%%%%%%%%%%%%%%%%%%%%%%%%%%%%%%%%%%%%%%%%%%%%%%%%%%%%%%%%%%%%%%%%%%%%%
\subsubsection*{Comment on Complementary Series Continuum}
If the spectrum extends into the complementary series, the leading soft
behavior is instead controlled by the lightest endpoint of the spectrum.
Suppose that the continuum starts at \(m=m_0<3H/2\), and define $\nu_0
\equiv
\sqrt{\frac94-\frac{m_0^2}{H^2}} $. After changing the integration variable from \(m^2\) to \(\nu\), the
complementary-series contribution can be written schematically as
\begin{align}
\mathcal F_{\rm comp}(r)
\sim
\int_0^{\nu_0}
\mathrm{d}\nu\,
W_{\rm comp}(\nu)\,
r^{-\nu},
\end{align}
where \(r\ll1\) denotes the relevant soft momentum ratio.  If the effective
spectral weight behaves near the endpoint as
\begin{align}
W_{\rm comp}(\nu)
\simeq
c_\beta(\nu_0-\nu)^\beta,
\qquad
\beta>-1,
\end{align}
the endpoint contribution is
\begin{align}
\mathcal F_{\rm comp}(r)
\sim
c_\beta\Gamma(\beta+1)
\frac{r^{-\nu_0}}
{|\log r|^{\beta+1}},
\qquad
r\to0 .
\label{comp_endpoint}
\end{align}
Thus the continuum gives the usual non-oscillatory power law associated with
the lightest complementary-series state, supplemented by a logarithmic
damping determined by the threshold behavior.  This should be distinguished
from the dephasing of principal-series clock oscillations discussed above.

%%%%%%%%%%%%%%%%%%%%%%%%%%%%%%%%%%
\subsection{Remarks on Analytic Background Contributions}\label{sec.BG}

Throughout the discussion above, we have focused on the non-analytic part of the seed
function, $\mathcal I_{\rm Signal}$.  This is the part that carries the cosmological-collider
clock signal.  In contrast, the analytic background contribution $\mathcal I_{\rm BG}$ in
Eq.~\eqref{Seed_dec} is analytic in momentum and is therefore sensitive
to the UV behavior of the spectral density.

This distinction becomes important for continuum spectra.  For a fixed principal-series
mass, the signal part is exponentially suppressed at large $\mu$ by the Boltzmann factor
$e^{-\pi\mu}$, so that its spectral integral is insensitive to the far UV for the examples
considered above.  The analytic background part does not enjoy this suppression.  To discuss the UV behavior, let
$\widehat{\mathcal I}_{\rm BG}(\mu;r_1,r_2)$ denote the fixed-mass background
contribution before performing the spectral integral.  For example, for
$(p_1,p_2)=(0,0)$, its large-$\mu$ behavior takes the schematic form
\begin{align}
  \widehat{\mathcal I}_{\rm BG}(\mu;r_1,r_2)
  \sim
  \frac{A_0(r_1,r_2)}{\mu^2}
  + \mathcal O(\mu^{-4}),
  \qquad \mu\gg 1 .
\end{align}
Therefore, for a power-law spectral density $\varrho(\mu)=\mu^\alpha$, the
large-$\mu$ behavior of the spectral integrand is
\begin{align}
  2\mu\,\varrho(\mu)\,
  \widehat{\mathcal I}_{\rm BG}(\mu;r_1,r_2)
  \sim
  \mu^{\alpha-1}.
\end{align}
The corresponding spectral integral is UV convergent for $\alpha<0$, logarithmically
divergent for $\alpha=0$, and power-law divergent for $\alpha>0$.

This UV sensitivity is not a universal cosmological-collider signal.  Rather,
it reflects the fact that analytic background terms are local contributions in
the low-energy effective theory.  Their finite parts must be specified together
with local counterterms, or equivalently by the UV completion.  For this reason,
in the following we focus on the non-analytic signal part when discussing
universal continuum effects.

%%%%%%%%%%%%%%%%%%%%%%%%%%%%%%%%%%%%%%%%%%%%%%%%%%%%%%%%%%%%%%%%%%%%%%%%%%%%%%%%%%%%%%%%%%%%%%%%%%%%%%%%%%%%%%%%%%%%%%%%%%%%%%%%%%%%%%%%%%%%%%%%%%%%%%%%%%%%%%%%%%%%%%%%%%%%%%%%%%%%%%%%%%%%%%%%%%%%%%%%%%%%%%%%%%%%%%%%%%%%%%%%%%%%%%%%
\section{UV Realization: RS2-like Inflationary Braneworld}
\label{UV}

So far, we have treated the continuous mass spectrum phenomenologically by
introducing a spectral function by hand.  In this section, we discuss a concrete
UV realization of such a continuum spectrum.  We consider a Randall--Sundrum II
(RS2) type setup~\cite{Randall:1999vf}, extended to an inflationary brane universe;
see, e.g., Ref.~\cite{Langlois:2002bb} for a review of brane cosmology.  In this
scenario, our four-dimensional inflating universe is identified with a brane
embedded in a five-dimensional anti-de Sitter bulk, and the inflaton is localized
on the brane. Related warped higher-dimensional inflationary scenarios involving a
brane-localized inflaton and dynamical radion evolution have recently been
studied in Ref.~\cite{Mishra:2025ofh}.

The key point for our purpose is that tensor perturbations in the five-dimensional
bulk contain, in addition to the localized four-dimensional graviton zero mode, a
continuum of Kaluza--Klein (KK) gravitons.  From the four-dimensional point
of view, this KK tower appears as a continuum of massive spin-2 states with a mass
gap of order the Hubble scale.  Therefore, it provides a natural UV realization of
the continuum spectra discussed in the previous section.

In contrast to the toy examples considered above, the spectral weight of the KK
continuum is not arbitrary.  It is determined by the extra-dimensional wavefunctions
and, in particular, by their values on the brane.  Since the inflaton is localized
on the brane, the same wavefunction factors also control the coupling of the KK
gravitons to inflaton fluctuations.  Our goal in this section is to derive this
KK spectral weight and to understand how the corresponding continuum modifies the
cosmological-collider signal.

%%%%%%%%%%%%%%%%%%%%%%%%%%%%%%%%%%%%%%%%%%%%%%%%%%%%%%%%%%%%%%%%%%
\subsection{Setup}

We consider the five-dimensional action
\begin{align}
S
=
\frac{1}{2\kappa_5^2}
\int \mathrm{d}^5x \sqrt{-g}\left(R-2\Lambda_5\right)
+
\int \mathrm{d}^5x \sqrt{-g}\,\delta(y)\,\mathcal{L}_{\rm brane},
\label{5D_action}
\end{align}
where $\kappa_5^2=1/M_5^3$.  We use five-dimensional coordinates
$x^M=(t,\mathbf{x}^i,y)$, where $t$ is the cosmic time, $\mathbf{x}^i$ $(i=1,2,3)$ are the
three spatial coordinates, and $y$ denotes the fifth direction.  The first term
in Eq.~\eqref{5D_action} is the bulk gravitational action with negative
cosmological constant $\Lambda_5<0$, while the second term describes a brane
localized at $y=0$.

We take the background metric to be
\begin{align}
\mathrm{d}s^2
=
n(y)^2
\left[
-\mathrm{d}t^2
+
a(t)^2 \mathrm{d}\mathbf{x}^2
\right]
+
\mathrm{d}y^2 ,
\label{BG_metric}
\end{align}
where $n(y)$ is the warp factor and $a(t)$ is the four-dimensional scale factor. We choose the normalization \(n(0)=1\), so that \(t\) coincides with the
cosmic time on the brane.
The five-dimensional Einstein equation is
\begin{align}
G_{MN}+\Lambda_5 g_{MN}
=
\kappa_5^2 T_{MN}.
\label{Eeq}
\end{align}
For the background metric~\eqref{BG_metric}, we take the brane energy-momentum
tensor to be
\begin{align}
T^M{}_{N}
=
\operatorname{diag}
\left(
-\rho_b,
p_b,
p_b,
p_b,
0
\right)
\delta(y),
\label{brane_TMN}
\end{align}
where $\rho_b$ and $p_b$ are the energy density and pressure localized on the
brane. Note that \(\rho_b\) includes the brane tension as well as the vacuum energy
localized on the brane. Then the independent components of the Einstein equation are
\cite{Langlois:2002bb}
\begin{align}
(0,0):\quad
&
\left(\frac{\dot a}{a}\right)^2
-
\left[
(n')^2+n n''
\right]
=
\frac{\Lambda_5}{3}n^2
+
\frac{\kappa_5^2\rho_b}{3}\delta(y),
\\
(i,i):\quad
&
\left(\frac{\dot a}{a}\right)^2
+
2\frac{\ddot a}{a}
-
3\left[
(n')^2+n n''
\right]
=
\Lambda_5 n^2
-
\kappa_5^2 p_b \delta(y),
\\
(5,5):\quad
&
\left(\frac{\dot a}{a}\right)^2
-
2(n')^2
+
\frac{\ddot a}{a}
=
\frac{\Lambda_5}{3}n^2 .
\end{align}
Here a dot denotes a derivative with respect to the cosmic time $t$, while a prime
denotes a derivative with respect to the extra-dimensional coordinate $y$.  This
prime should not be confused with the conformal-time derivative used in the
previous section.

For simplicity, we now look for a solution in which the four-dimensional spacetime
is exactly de Sitter,
\begin{align}
a(t)=e^{Ht},
\qquad
H=\mathrm{const.}
\end{align}
Although deviations from exact de Sitter are necessary for realistic inflationary
cosmology and phenomenology, this approximation is sufficient for our present
purpose.  Under this assumption, the above equations reduce to
\begin{align}
(0,0):\quad
&
H^2-(n')^2-n n''
=
\frac{\Lambda_5}{3}n^2
+
\frac{\kappa_5^2\rho_b}{3}\delta(y),
\label{00}
\\
(i,i):\quad
&
H^2-(n')^2-n n''
=
\frac{\Lambda_5}{3}n^2
-
\frac{\kappa_5^2 p_b}{3}\delta(y),
\\
(5,5):\quad
&
H^2-(n')^2
=
\frac{\Lambda_5}{6}n^2 .
\label{55}
\end{align}
The first two equations are equivalent if the brane source satisfies
$p_b=-\rho_b$.  This corresponds to a brane-localized vacuum energy sourcing the
four-dimensional de Sitter expansion, and we assume this equation of state in the
following.  The remaining equations then determine the warp factor $n(y)$ and the
Hubble scale $H$.

To solve the Einstein equations, we also need the junction conditions at the brane.
They are given by~\cite{Langlois:2002bb}
\begin{align}
\left(\frac{n'}{n}\right)_{0^+}
=
-\frac{\kappa_5^2}{6}\rho_b,
\qquad
\left(\frac{n'}{n}\right)_{0^+}
=
\frac{\kappa_5^2}{6}
\left(3p_b+2\rho_b\right).
\label{JC}
\end{align}
Again, the two conditions are equivalent for $p_b=-\rho_b$.

%%%%%%%%%%%%%%%%%%%%%%%%%%%%%%%%%%%%%%%%%%%%%%%%%%%%%%%%%%%%%%%%%%
\subsection{Background Solutions}

We now solve the background Einstein equations derived above.  Combining
Eqs.~\eqref{00} and~\eqref{55}, and neglecting the delta-function contribution
away from the brane, we obtain
\begin{align}
n''=K^2 n,
\qquad
K^2\equiv -\frac{\Lambda_5}{6}.
\end{align}
Thus, on the $y>0$ side, the bulk solution is
\begin{align}
n(y)=c_1 e^{Ky}+c_2 e^{-Ky}.
\label{sol_n}
\end{align}
Since the brane is located at $y=0$ and we impose the $Z_2$ symmetry, the solution
on the full bulk is obtained by replacing $y$ with $|y|$.  In the following,
unless otherwise stated, we work on the $y\geq 0$ side and omit the absolute value
for notational simplicity.

The constants $c_1$ and $c_2$ are fixed by the normalization condition
\begin{align}
n(0)=1,
\qquad
\Longleftrightarrow
\qquad
c_1+c_2=1,
\end{align}
and by the junction condition~\eqref{JC},
\begin{align}
\frac{n'(0)}{n(0)}
=
-\frac{\kappa_5^2}{6}\rho_b,
\qquad
\Longleftrightarrow
\qquad
\frac{c_1-c_2}{c_1+c_2}
=
-\frac{\kappa_5^2}{6K}\rho_b.
\end{align}

We decompose the brane energy density and pressure as
\begin{align}
\rho_b=\sigma_0+\rho_0,
\qquad
p_b=-\sigma_0+p_0,
\end{align}
where $\sigma_0$ is the brane tension.  In the usual static RS2 model, the brane
tension is fine-tuned as
\begin{align}
\sigma_0=\frac{6K}{\kappa_5^2}.
\end{align}
The quantities $\rho_0$ and $p_0$ denote the energy density and pressure of the
brane-localized matter sector.  Equivalently, the brane Lagrangian is written as
\begin{align}
\mathcal{L}_{\rm brane}
=
\mathcal{L}_{\rm matter}
-
\sigma_0 .
\end{align}
In the following, we assume that this matter sector is dominated by a brane-localized
inflaton, so that $p_0= -\rho_0$.  This is consistent with the de Sitter
condition $p_b=-\rho_b$ imposed above.

The constants in Eq.~\eqref{sol_n} are then
\begin{align}
c_1
=
-\frac{\kappa_5^2}{12K}\rho_0
\equiv
-\frac{\alpha}{2},
\qquad
c_2
=
1+\frac{\kappa_5^2}{12K}\rho_0
\equiv
1+\frac{\alpha}{2},
\end{align}
where we have introduced
\begin{align}
\alpha
\equiv
\frac{\kappa_5^2\rho_0}{6K}
=
\frac{\rho_0}{\sigma_0}.
\end{align}
The parameter $\alpha$ characterizes the deviation from the usual static RS2
background.  The warp factor can then be written as
\begin{align}
n(y)
=
\cosh(Ky)
-
(1+\alpha)\sinh(Ky),
\label{sol_n_y}
\end{align}
for $y\geq 0$.

Substituting this solution into Eq.~\eqref{55}, the Hubble scale is determined as
\begin{align}
H^2
=
K^2\alpha(2+\alpha)
=
\frac{\kappa_5^2 K}{3}\rho_0
\left(
1+\frac{\rho_0}{2\sigma_0}
\right).
\label{H_alpha}
\end{align}
The term quadratic in $\rho_0$ is the characteristic high-energy correction in
brane cosmology; see Refs.~\cite{Binetruy:1999hy,Maartens:1999hf} for more
details.

It is sometimes convenient to introduce the conformal coordinate
\begin{align}
z
=
\int_0^y \frac{\mathrm{d}y'}{n(y')},
\end{align}
such that the metric~\eqref{BG_metric} takes the conformally flat form
\begin{align}
\mathrm{d}s^2
=
e^{-A(z)}
\left[
-\mathrm{d}t^2
+
a(t)^2 \mathrm{d}\mathbf{x}^2
+
\mathrm{d}z^2
\right],
\label{BG_metric_2}
\end{align}
with $n(y(z))=e^{-A(z)/2}$. For the solution~\eqref{sol_n_y}, one finds
\begin{align}
e^{-A(z)/2}
=
\frac{H}{K}
\operatorname{csch}\left(H|z|+c\right),
\label{sol_A}
\end{align}
where
\begin{align}
c
=
2\tanh^{-1}
\left(
\sqrt{\frac{\alpha}{\alpha+2}}
\right).
\end{align}
The absolute value makes the \(Z_2\) symmetry manifest.  Using
\(H^2=K^2\alpha(2+\alpha)\), one has
\begin{align}
\sinh c=\frac{H}{K},
\qquad
\cosh c=1+\alpha ,
\end{align}
so that the normalization \(e^{-A(0)/2}=n(0)=1\) is automatically satisfied.

In the static limit $\alpha\to0$, and hence $H\to0$, we recover
\begin{align}
n(y)=e^{-K|y|},
\qquad {\rm{or}}\qquad
e^{-A(z)/2}
=
\frac{1}{1+K|z|},
\end{align}
which is the usual RS2 warp factor~\cite{Csaki:2000fc}.

Before moving on to the KK spectrum, let us comment on the effective four-dimensional Planck mass. It is obtained by dimensional
reduction of the five-dimensional Einstein--Hilbert term.  Including the two
$Z_2$-symmetric sides of the bulk, it is given by
\begin{align}
M_{\rm Pl}^2
=
2M_5^3
\int_0^\infty \mathrm{d}z\, e^{-3A(z)/2}.
\end{align}
Using Eq.~\eqref{sol_A}, this becomes
\begin{align}
M_{\rm Pl}^2
=
\frac{M_5^3}{K}
\left[
1+\alpha
+
\frac{\alpha(2+\alpha)}{2}
\log\frac{\alpha}{\alpha+2}
\right].
\label{Mpl_alpha}
\end{align}
In the static limit $\alpha\to0$, this reduces to the usual RS2 relation
\begin{align}
M_{\rm Pl}^2
\to
\frac{M_5^3}{K}
.
\end{align}
The numerical examples below use \(\alpha\ll1\), corresponding to the low-energy
regime \(\rho_0\ll\sigma_0\) and \(H/K\ll1\).

%%%%%%%%%%%%%%%%%%%%%%%%%%%%%%%%%%%%%%%%%%%%%%%%%%%%%%%%%%%%%%%%%%
\subsection{KK Graviton Spectrum}

We now discuss tensor fluctuations around the background~\eqref{BG_metric_2}.
We write the perturbed metric as
\begin{align}
\mathrm{d}s^2
=
e^{-A(z)}
\left[
-\mathrm{d}t^2
+
a(t)^2\mathrm{d}\mathbf{x}^2
+
\mathrm{d}z^2
+
h_{ij}(t,\mathbf{x},z)\mathrm{d}x^i\mathrm{d}x^j
\right].
\label{h_fluctuation}
\end{align}
Here $h_{ij}$ denotes the spatial metric perturbation with lower indices.\footnote{If one
instead introduces the tensor perturbation $\gamma_{ij}$ by
$g_{ij}=e^{-A}a^2(\delta_{ij}+\gamma_{ij})$, the two conventions are related by
$h_{ij}=a^2\gamma_{ij}$.}
In general, additional scalar and vector fluctuations may also be present.
In the following, however, we restrict our attention to the tensor sector
and impose the transverse-traceless conditions
\begin{align}
\partial^i h_{ij}=0,
\qquad
\delta^{ij}h_{ij}=0.\label{TT}
\end{align}

Then, the EOM for the tensor mode $h_{ij}$ is given by (see Ref.~\cite{Kumar:2018jxz, Kumar:2025anx} and Appendix~\ref{G_EOM} for details)
\begin{align}
-\Box_{\rm{dS}} h_{i j}+2 H^2 h_{i j}+\frac{3}{2} A^{\prime} h^{ \prime}_{i j}-h^{\prime \prime}_{ij}=0,\label{gravi_eom_RS2}
\end{align}
where $\Box_{\rm dS}$ denotes the covariant d'Alembertian on the
four-dimensional de Sitter background, $
\mathrm{d}\bar s_4^2
=
-\mathrm{d}t^2+a(t)^2\mathrm{d}\mathbf{x}^2$.
Explicitly, for the tensor perturbation considered here, it gives
\begin{align}
-\Box_{\rm dS} h_{ij}
=
\ddot h_{ij}
-
H\dot h_{ij}
-
a^{-2}\partial^2 h_{ij}
-
4H^2h_{ij},
\end{align}
where $\partial^2\equiv \delta^{ij}\partial_i\partial_j$.
Defining
\begin{align}
\tilde h_{ij}=e^{-3A/4}h_{ij},
\end{align}
the equation can be rewritten as
\begin{align}
-\Box_{\rm{dS}}\tilde{h}_{i j}+2 H^2 \tilde{h}_{i j}+\left[\frac{9}{16}\left(A^{\prime}\right)^2-\frac{3}{4} A^{\prime \prime}\right] \tilde{h}_{i j}-\tilde{h}^{\prime \prime}_{ij}=0. \label{EOM_htil}
\end{align}

Now we perform a KK decomposition of the tensor perturbation as
\begin{align}
\tilde{h}_{i j}(t, \mathbf{x}, z)=\hat{h}_{i j}(t, \mathbf{x}) \psi(z),
\end{align}
where $\hat{h}_{i j}$ describes a four-dimensional tensor mode with KK mass \(m\) and satisfies
\begin{align}
-\Box_{\rm{dS}}\hat{h}_{i j}+2H^2\hat{h}_{i j}=-m^2\hat{h}_{i j}.\label{KKdec}
\end{align}
Substituting this decomposition into Eq.~\eqref{EOM_htil}, we obtain the following Schrödinger-type equation for the extra-dimensional wavefunction \(\psi(z)\):
\begin{align}
-\psi^{\prime \prime}(z)+\left(\frac{9}{16} A^{\prime 2}-\frac{3}{4} A^{\prime \prime}\right) \psi(z)=m^2 \psi(z).    \label{SE}
\end{align}
This equation can be interpreted as a Schrödinger equation with the effective potential
\begin{align}
\nonumber V(z)&=  \frac{9}{16} A^{\prime 2}-\frac{3}{4} A^{\prime \prime}\\
&=  \frac{9}{4} H^2+\frac{15}{4} H^2 \operatorname{csch}^2(H |z|+c)-3H\coth(c)\delta(z),
\end{align}
where we have substituted the background solution $A(z)$ given in Eq.~\eqref{sol_A} and the brane-localized \(\delta\)-function term originates from the second derivative of \(|z|\) contained in \(A''\). In the limit with $\alpha\rightarrow 0$, it is reduced to the usual RS2 effective potential $15K^2/4\times (1+K|z|)^{-2}-3 K \delta(z)$.

Since the Schrödinger potential asymptotes to the constant
$9H^2/4$ as $z\rightarrow\infty$, the continuum KK modes satisfy
$m^2>9H^2/4$, implying that the continuum spectrum starts at
$m=3H/2$. On the other hand, the attractive delta-function term localized at the brane suggests
the existence of a normalizable bound state separated from the
continuum spectrum. The unique normalizable bound state corresponds to the
four-dimensional massless graviton.

Integrating the Schrödinger equation across the brane and using the $Z_2$ symmetry,
we obtain
\begin{align}
\psi'(0^+)
=
-\frac{3H}{2}\coth c\,\psi(0).\label{BD_psi}
\end{align}
In the following, we solve the bulk equation~\eqref{SE} on the half-line
\(z\ge0\), with the effect of the brane-localized
\(\delta\)-function encoded in the boundary condition
\eqref{BD_psi}.

\subsubsection*{Zero Mode}
For \(m=0\), Eq.~\eqref{SE} admits the solution
\begin{align}
\psi_0(z)=\mathcal{N}_0 \operatorname{csch}^{3 / 2}(H z+c),
\end{align}
which satisfies both Eq.~\eqref{SE} and the boundary condition
\eqref{BD_psi}. The normalization constant $\mathcal{N}_0 $ is fixed by 
\begin{align}
\int_0^{\infty} \mathrm{d} z\left|\psi_0(z)\right|^2=1, \quad \Rightarrow \quad \mathcal{N}_0=\sqrt{\frac{2 H}{\operatorname{coth} c \operatorname{csch} c+\ln \tanh (c / 2)}}. \label{N_0}
\end{align}
When $\alpha \rightarrow 0$, we recover the zero mode wave function for the usual RS2 model: $\psi^{\rm{RS2}}_0(z)=\sqrt{2K}(1+Kz)^{-3/2} $.

On the other hand, the four-dimensional tensor mode satisfies Eq.~\eqref{KKdec} with $m=0$ and its positive-frequency solution is given by
\begin{align}
\hat h^{(0)}_{ij}(\tau,\mathbf{x})
=
\sum_{\lambda=\pm2}
\int \frac{\mathrm{d}^3k}{(2\pi)^3}
e^{i\mathbf{k}\cdot\mathbf{x}}
h_\lambda^{(0)}(\tau,k)
e_{ij}^{(\lambda)}(\hat{\mathbf{k}}),
\end{align}
where we switch to the conformal time $\tau$ and the mode function in the Bunch-Davies vacuum is given by
\begin{align}
h_\lambda^{(0)}(\tau,k)
=
a^2(\tau)\times \frac{H}{\sqrt{2k^3}}
(1+ik\tau)e^{-ik\tau},\label{G_0}
\end{align}
with $a(\tau)=-1/(H\tau)$, and $e_{ij}^{(\lambda)}$ is the polarization tensors normalized by $e_{i j}^{(\lambda)} e_{i j}^{\left(\lambda^{\prime}\right) *}=2 \delta_{\lambda \lambda^{\prime}}$ and $e_{i j}^{(-\lambda)}=e_{i j}^{(\lambda) *}$.

Combining the extra-dimensional wavefunction and the four-dimensional tensor
mode, the zero-mode contribution to the rescaled perturbation is
\begin{align}
\tilde h_{ij}^{(0)}
=
\psi_0(z)\hat h_{ij}^{(0)}(\tau,\mathbf{x}).
\end{align}
The corresponding perturbation in the original variable is obtained from
\(h_{ij}=e^{3A/4}\tilde h_{ij}\).
%%%%%%%%%%%%%%%%%%%%%%%%%%%%%%%%%%%%%%%%%%%%
\subsubsection*{Massive Continuum Modes}
Let us move to the continuous mode with $m>3H/2$. The two independent solutions of Eq.~\eqref{SE} are given by 
\begin{align}
f_\mu^\pm (z)= e^{\mp \pi \mu/2}\Gamma(1\mp i \mu) P_{3 / 2}^{\pm i \mu}\left[\operatorname{coth} (H z+c)\right],\label{f_mu}
\end{align}
where $P_\ell^m(z)$ denotes the associated Legendre function and
\begin{align}
 \mu \equiv \sqrt{\left(\frac{m}{H}\right)^2-\frac{9}{4}}>0.\label{def_mu}
\end{align}
Their asymptotic behavior is  
\begin{align}
 f_\mu^\pm (z)   \sim e^{\pm i \mu (H z+c)}.\label{asymp_mode}
\end{align}
The boundary condition~\eqref{BD_psi} fixes the relative coefficient between
$f_\mu^\pm$, yielding
\begin{align}
\psi_m(z)=\mathcal{N}_m\left[e^{-i\delta_\mu}f_\mu^+ (z)+e^{i\delta_\mu} f_\mu^- (z)\right]  \label{sol_psi_m}  
\end{align}
where 
\begin{align}
e^{2i\delta_\mu}\equiv -\frac{\beta_\mu}{\beta_\mu^*}, \quad \beta_\mu\equiv \left(\frac{3}{2}+i \mu\right) \Gamma(1-i \mu) P_{1 / 2}^{i \mu}(\operatorname{coth} c).  \label{R_mu}  
\end{align}
$\mathcal N_m$ is a normalization constant determined by
\begin{align}
\int_0^{\infty} \mathrm{d} z\ \psi_m^*(z) \psi_{m^{\prime}}(z)=\delta\left(m-m^{\prime}\right),    
\end{align}
and using the asymptotic form of the mode functions~\eqref{asymp_mode},
the normalization condition yields 
\begin{align}
 \mathcal{N}_m=   \sqrt{\frac{m/H}{2 \pi \mu}}.\label{N_m}
\end{align}

\begin{figure}[H]
 \begin{minipage}{0.5\hsize}
  \begin{center}
   \includegraphics[width=80mm]{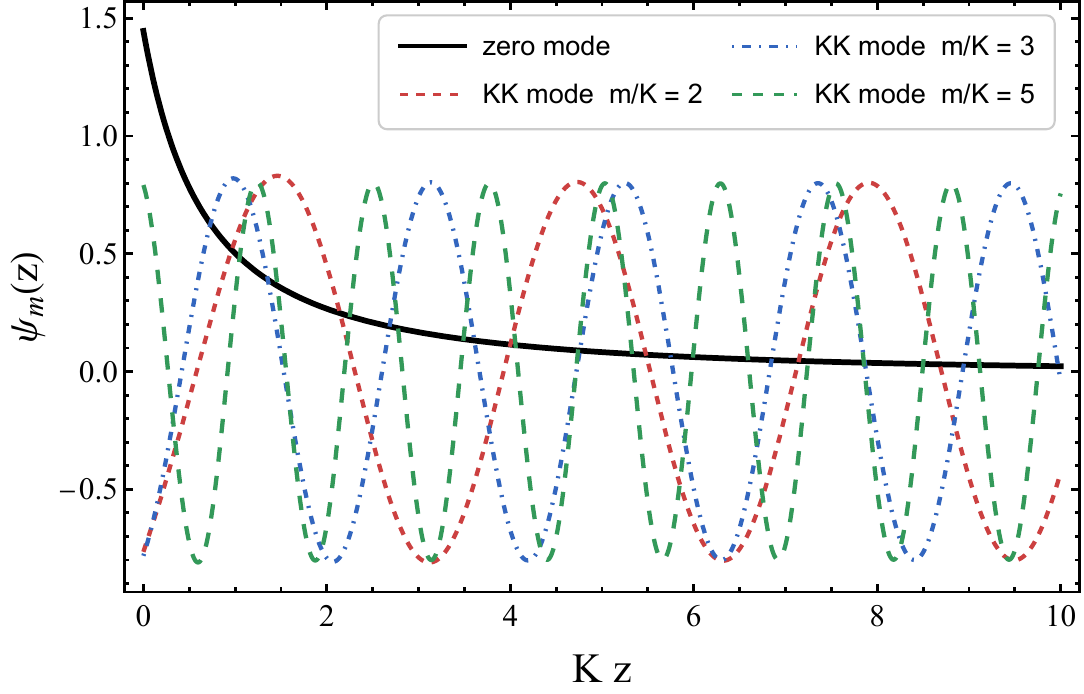}
  \end{center}
 \end{minipage}
 \begin{minipage}{0.5\hsize}
  \begin{center}
   \includegraphics[width=80mm]{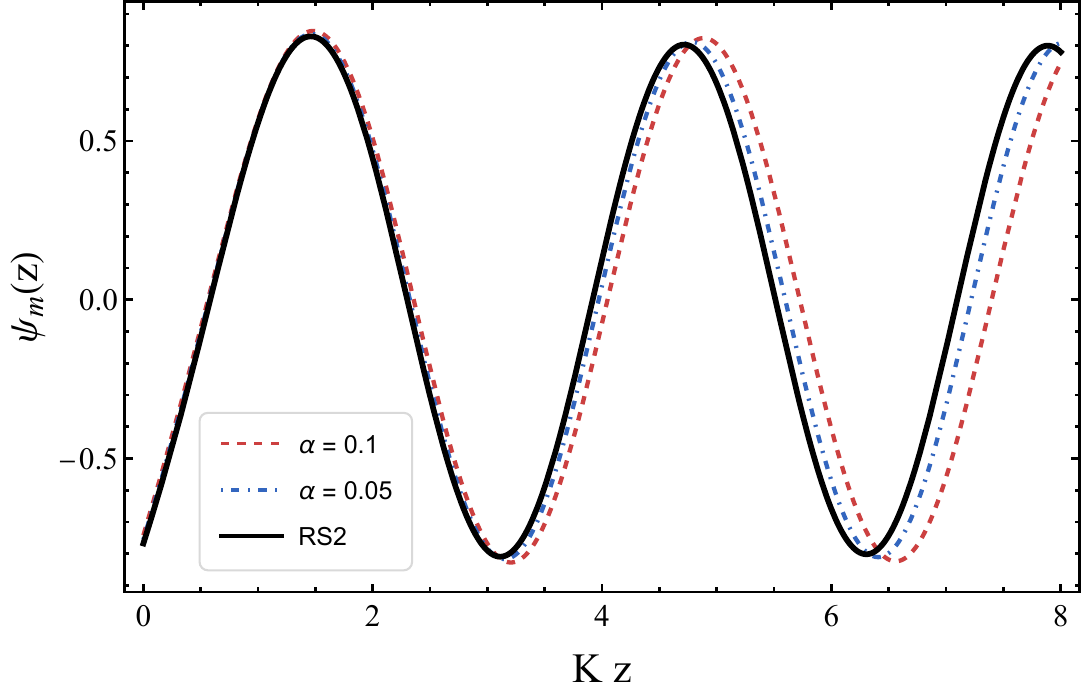}
  \end{center}
 \end{minipage}
   \caption{
Wavefunctions of the localized graviton zero mode and the continuum KK modes.
\textit{Left panel}: The zero mode and several continuum modes for $\alpha=0.01$,
plotted as functions of $Kz$.  We set $K=1$ and show modes with
$m/K=2,3,5$.  For this value of $\alpha$, these correspond to
$m/H\simeq 14.1,21.2,35.3$, respectively.
\textit{Right panel}: Comparison with the static RS2 continuum wavefunction,
with $m/K=2$ fixed.  The de Sitter-brane wavefunctions approach the static RS2
wavefunction as $\alpha$ decreases. }
\label{FigKK}
\end{figure}

In the left panel of Fig.~\ref{FigKK}, we plot the wave functions of the
zero mode, $\psi_0(z)$, together with several continuum KK modes,
$\psi_m(z)$. One can see that the zero mode is localized near the brane,
whereas the massive modes extend into the bulk and exhibit increasingly
rapid oscillations as $m/K$ increases. 
We also confirmed that the solution~\eqref{sol_psi_m} reduces to the
well-known RS2 continuum KK mode function~\cite{Randall:1999vf}. In the limit $\alpha\to0$,
the associated Legendre functions appearing in Eq.~\eqref{sol_psi_m}
reduce to combinations of Bessel functions,
\begin{align}
\psi^{\rm{RS2}}_m(z) = \frac{\sqrt{\frac{m}{K}(1+K z)}}{\sqrt{J_1^2\left(\frac{m}{K}\right)+Y_1^2\left(\frac{m}{K}\right)}} \left[-Y_1\left(\frac{m}{K}\right) J_2\left(\frac{m}{K}(1+K z)\right)+J_1\left(\frac{m}{K}\right) Y_2\left(\frac{m}{K}(1+K z)\right)\right],
\end{align}
where $J_n(x)$ and $Y_n(x)$ denote the Bessel functions of the first
and second kinds, respectively. The right panel of Fig.~\ref{FigKK} shows a numerical comparison.

We now specify the four-dimensional helicity-$\pm2$ mode functions of the massive
KK gravitons.  The four-dimensional tensor perturbation satisfies
Eq.~\eqref{KKdec}.
We focus on the helicity-$\pm2$ tensor modes, while the helicity-$0$ and
helicity-$\pm1$ components of the massive spin-2 field are not considered here.
See Ref.~\cite{Tong:2022cdz} for a complete treatment.
Accordingly, we write
\begin{align}
\hat{h}_{i j}^{(m)}(\tau, \mathbf{x})=\sum_{\lambda= \pm 2} \int \frac{\mathrm{d}^3 k}{(2 \pi)^3} e^{i \mathbf{k} \cdot \mathbf{x}} h_\lambda^{(m)}(\tau, k) e_{ij}^{(\lambda)}(\hat{\mathbf{k}})+\cdots    
\end{align}
where the ellipsis denotes the helicity-$0$ and helicity-$\pm1$ components. Substituting the helicity decomposition into Eq.~\eqref{KKdec},
we find that the helicity $\lambda=\pm2$ mode functions satisfy
\begin{align}
\left(\partial_\tau^2+\frac{2}{\tau}\partial_\tau +k^2+\frac{m^2/H^2-2}{\tau^2}\right)h_{\lambda}^{(m)} =0.
\end{align}
The Bunch-Davies solution is given by
\begin{align}
h^{(m)}_{\lambda} (\tau,k) =
e^{-\frac{\pi}{2}\mu+i\frac{\pi}{4}}
\frac{\sqrt{\pi }}{2H}(-\tau)^{-1/ 2}
H_{i\mu}^{(1)}(-k \tau), \label{G_m} 
\end{align}
where $\mu$ is defined in Eq.~\eqref{def_mu}.

%%%%%%%%%%%%%%%%%%%%%%%%%%%%%%%%%%%%%%%%%%%
\subsubsection*{Summary of KK Solutions}

Within the tensor sector considered here, the rescaled perturbation is
decomposed into the localized massless graviton and the continuum KK modes as
\begin{align}
\tilde h_{ij}
=
\psi_0(z)\hat h^{(0)}_{ij}(\tau,\mathbf{x})
+
\int_{3H/2}^{\infty}
\mathrm{d}m\,
\psi_m(z)\hat h^{(m)}_{ij}(\tau,\mathbf{x}) .
\label{sol_KKtower}
\end{align}
Here \(\psi_0\) denotes the normalizable zero-mode wavefunction,
\begin{align}
\psi_0(z)
=
\mathcal{N}_0
\operatorname{csch}^{3/2}(H z+c),
\qquad
z\ge0,
\end{align}
with \(\mathcal{N}_0\) given in Eq.~\eqref{N_0}.  The continuum KK
wavefunctions are
\begin{align}
\psi_m(z)
=
\mathcal{N}_m
\left[
e^{-i\delta_\mu}f_\mu^+(z)
+
e^{i\delta_\mu}f_\mu^-(z)
\right],
\end{align}
where
\begin{align}
\mu
=
\sqrt{\frac{m^2}{H^2}-\frac94}.
\end{align}
The normalization constant \(\mathcal N_m\), the phase shift \(\delta_\mu\),
and the basis functions \(f_\mu^\pm\) are given in
Eqs.~\eqref{N_m}, \eqref{R_mu}, and \eqref{f_mu}, respectively.  Thus the
spectrum consists of a localized massless graviton at \(m=0\) and a continuum
of KK gravitons starting at \(m=3H/2\), separated from the zero mode by a mass
gap \(3H/2\).

The four-dimensional tensor modes appearing in Eq.~\eqref{sol_KKtower} are
written as
\begin{align}
\hat h^{(0)}_{ij}(\tau, \mathbf{x})
&=
\sum_{\lambda=\pm2}
\int\frac{\mathrm{d}^3k}{(2\pi)^3}
e^{i\mathbf k\cdot\mathbf x}
h_\lambda^{(0)}(\tau, k)
e_{ij}^{(\lambda)}(\hat{\mathbf{k}}),
\\
\hat h^{(m)}_{ij}(\tau, \mathbf{x})
&=
\sum_{\lambda=\pm2}
\int\frac{\mathrm{d}^3k}{(2\pi)^3}
e^{i\mathbf k\cdot\mathbf x}
h_\lambda^{(m)}(\tau, k)
e_{ij}^{(\lambda)}(\hat{\mathbf{k}}).
\end{align}
The explicit forms of the mode functions are given in
Eqs.~\eqref{G_0} and~\eqref{G_m}.  The helicity-0 and helicity-\(\pm1\)
components of the massive spin-2 field are not included in the present tensor
sector analysis.

%%%%%%%%%%%%%%%%%%%%%%%%%%%%%%%%%%%%%%%%%%%%%%%%%%%%%

\subsection{Quantization and KK Spectral Weight}\label{Quantization and KK Spectral Weight}

\begin{figure}[H]
  \begin{center}
   \includegraphics[width=100mm]{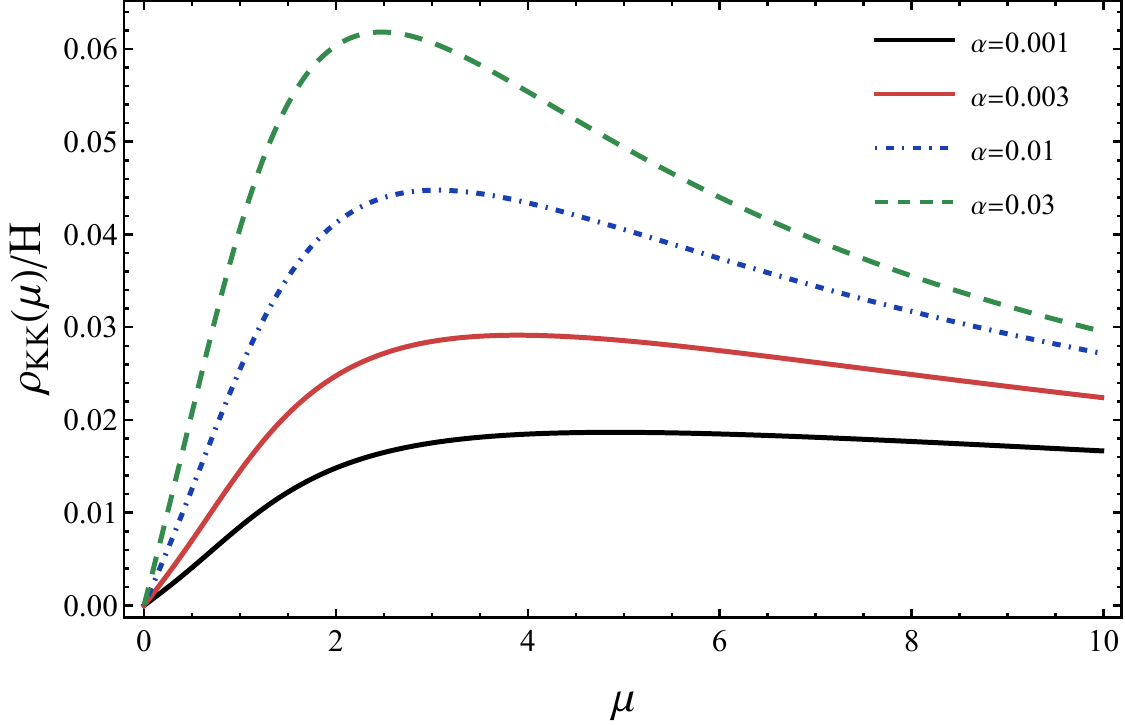}
  \end{center}
\caption{
KK spectral weight on the brane, plotted as a function of
\(\mu=\sqrt{m^2/H^2-9/4}\).  We show the dimensionless combination
\(\rho_{\rm KK}(\mu)/H\). The spectral weight vanishes linearly at the threshold \(\mu=0\), reaches a
maximum at an intermediate value of \(\mu\), and decreases as \(1/\mu\) in the
large-mass regime. The curves correspond to
\(\alpha=0.001, 0.003, 0.01,\) and \(0.03\).
}
\label{FigrhoKK}
\end{figure}

We now evaluate the tensor perturbation on the brane and identify the effective
spectral weight of the KK continuum. Since \(A(0)=0\), the rescaled perturbation \(\tilde h_{ij}\) coincides with
the original metric perturbation \(h_{ij}\) on the brane.  Therefore, from
Eq.~\eqref{sol_KKtower}, the brane value relevant for brane-localized matter is 
\begin{align}
\tilde h_{ij}(\tau,\mathbf{x},0)
=
\psi_0(0)\hat h^{(0)}_{ij}(\tau,\mathbf{x})
+
\int_{3H/2}^{\infty}
\mathrm{d}m\,
\psi_m(0)\hat h^{(m)}_{ij}(\tau,\mathbf{x}) .
\label{KK_0}
\end{align}
For the helicity-$\pm2$ part of the massive KK tower, we quantize
\begin{align}
\hat h^{(m)}_{ij}(\tau,\mathbf{x})
=
\sum_{\lambda=\pm2}
\int\frac{\mathrm{d}^3k}{(2\pi)^3}
\left[
e_{ij}^{(\lambda)}(\hat{\mathbf{k}})
h_{k,m}(\tau)
b_{\mathbf{k},\lambda,m}
e^{i\mathbf{k}\cdot\mathbf{x}}
+
e_{ij}^{(\lambda)*}(\hat{\mathbf{k}})
h^*_{k,m}(\tau)
b^\dagger_{\mathbf{k},\lambda,m}
e^{-i\mathbf{k}\cdot\mathbf{x}}
\right],
\label{h_m_quant}
\end{align}
and similarly for the zero mode.  The annihilation and creation operators satisfy
\begin{align}
\left[
b_{\mathbf{k},\lambda,m},
b^\dagger_{\mathbf{k}',\lambda',m'}
\right]
=
(2\pi)^3
\delta_{\lambda\lambda'}
\delta^{(3)}(\mathbf{k}-\mathbf{k}')
\delta(m-m') .
\label{CR_KK_m}
\end{align}
With this normalization, the continuum part of the brane two-point function
contains the measure
\begin{align}
\int_{3H/2}^{\infty}
\mathrm{d}m\,
|\psi_m(0)|^2
=
\int_{(3H/2)^2}^{\infty}
\frac{\mathrm{d}m^2}{H^2}\,
\rho_{\rm KK}(m^2),
\end{align}
where the second equality is obtained by changing the integration variable from \(m\) to \(m^2\), and we define the KK spectral weight by
\begin{align}
\rho_{\rm KK}(m^2)
\equiv
\frac{H^2}{2m}
|\psi_m(0)|^2 .
\label{rho_m}
\end{align}
With this convention, $\rho_{\rm KK}$ has mass dimension one.  This is because
the continuum modes are normalized as
$\int_0^\infty \mathrm{d}z\,\psi_m^*(z)\psi_{m'}(z)=\delta(m-m')$.
To compare with the dimensionless spectral function used in Sec.~\ref{sec:Conti},
we will often use the dimensionless combination \(\rho_{\rm KK}/H\), which is
also the quantity shown in the plots below.
In Fig.~\ref{FigrhoKK}, we plot the dimensionless quantity
\(\rho_{\rm KK}(\mu)/H\) as a function of \(\mu=\sqrt{m^2/H^2-9/4}\).  The continuum starts at
\(\mu=0\), corresponding to the mass gap \(m=3H/2\).

The asymptotic behavior of the KK spectral weight can be understood directly
from the brane value of the continuum wavefunction.  Near the threshold, \(\mu=0\), the continuum wavefunction on the brane is
suppressed.  Expanding the exact solution~\eqref{sol_psi_m} around
\(\mu=0\), one finds
\begin{align}
|\psi_{m(\mu)}(0)|^2
\propto
\mu ,
\qquad
\mu\ll1 .
\end{align}
Therefore,
\begin{align}
\frac{\rho_{\rm KK}(\mu)}{H}
=
\frac{
|\psi_{m(\mu)}(0)|^2
}{
2\sqrt{\mu^2+\frac94}
}\propto
\mu ,
\qquad
\mu\ll1 .\label{rho_KK_asmp}
\end{align}

In the opposite limit, the oscillation scale of the continuum wavefunction in
the extra dimension becomes much shorter than the scale over which the potential
varies.  The massive mode then approaches a freely normalized standing wave on
the half-line.  With the normalization
\begin{align}
\int_0^\infty \mathrm{d}z\,
\psi_m^*(z)\psi_{m'}(z)
=
\delta(m-m'),
\end{align}
this gives
\begin{align}
|\psi_m(0)|^2
\to
\frac{2}{\pi},
\qquad
m/K\gg1 .
\end{align}
Hence
\begin{align}
\frac{\rho_{\rm KK}(\mu)}{H}
\simeq
\frac{1}{\pi\sqrt{\mu^2+\frac94}}
\sim
\frac{1}{\pi\mu},
\qquad
m/K\gg1 .
\end{align}
Thus the dimensionless KK spectral weight grows linearly from the threshold,
reaches a maximum at an intermediate value of \(\mu\), and decreases in the large-mass regime \(m/K\gg1\).

The Schwinger--Keldysh propagators for the brane tensor perturbation can now be
constructed directly.  The Wightman propagator is\footnote{Here and below,
\(D^{\mathrm{a}\mathrm{b}}_{ij,k\ell}\) denotes the brane-to-brane
propagator with the overall gravitational coupling stripped off.  With our
normalization, the two-point function of the metric perturbation is
\(\kappa_5^2D^{\mathrm{a}\mathrm{b}}_{ij,k\ell}\). In addition, with the \(\delta(m-m')\)
normalization of the continuum modes, the relevant mass dimensions are
\([\psi_0]=1/2\), \([\psi_m]=0\), \([\rho_{\rm KK}]=1\), and
\([h_{k,0}]=[h_{k,m}]=-1/2\).  Thus the reduced propagator is dimensionless.}
\begin{align}
D^{-+}_{ij,k\ell}(\mathbf{k};\tau_1,\tau_2)
=
\Pi_{ij,k\ell}(\hat{\mathbf{k}})
\left[
D^{(0)}_{-+}(k;\tau_1,\tau_2)
+
D_{-+}^{\rm KK}(k;\tau_1,\tau_2)
\right],
\label{SK_KK}
\end{align}
where
\begin{align}
\Pi_{ij,k\ell}(\hat{\mathbf{k}})
\equiv
\sum_{\lambda=\pm2}
e_{ij}^{(\lambda)}(\hat{\mathbf{k}})
e_{k\ell}^{(\lambda)*}(\hat{\mathbf{k}})
\label{TT_projector}
\end{align}
is the helicity-\(\pm2\) polarization sum, namely the three-dimensional TT
projector, and
\begin{align}
D_{-+}^{(0)}(k;\tau_1,\tau_2)
&\equiv
\psi_0^2(0)
h_{k,0}(\tau_1)
h^*_{k,0}(\tau_2),
\\
D_{-+}^{\rm KK}(k;\tau_1,\tau_2)
&\equiv
\int_{(3H/2)^2}^{\infty}
\frac{\mathrm{d}m^2}{H^2}\,
\rho_{\rm KK}(m^2)
h_{k,m}(\tau_1)
h_{k,m}^*(\tau_2)
\label{D_KK_minusplus}
\end{align}
are the Wightman propagators for the zero mode and the continuum KK gravitons,
respectively.  Since the helicity \(+2\) and \(-2\) modes obey the same mode
equation in the parity-invariant de Sitter background, their mode functions are
identical, as in Eq.~\eqref{G_m}.  Therefore the polarization dependence
factorizes from the time-dependent part of the propagator, and we suppress the
helicity label of the mode functions.  The propagator \(D_{-+}^{\rm KK}\)
defined in this way is the helicity-\(\pm2\) counterpart of the scalar continuum
propagator discussed in Sec.~\ref{sec:Conti}.  Apart from the helicity-\(\pm2\) polarization projector, the main differences
are the tensor mode function and the fact that the spectral weight is fixed by
the underlying five-dimensional geometry.

The remaining SK propagators are defined as
\begin{align}
D^{+-}_{ij,k\ell}(\mathbf{k};\tau_1,\tau_2)
&=
D^{-+}_{k\ell,ij}(\mathbf{k};\tau_2,\tau_1)
=
\left[
D^{-+}_{ij,k\ell}(\mathbf{k};\tau_1,\tau_2)
\right]^*,
\label{SK_+-KK}
\\
D^{\pm\pm}_{ij,k\ell}(\mathbf{k};\tau_1,\tau_2)
&=
D^{\mp\pm}_{ij,k\ell}(\mathbf{k};\tau_1,\tau_2)
\theta(\tau_1-\tau_2)
+
D^{\pm\mp}_{ij,k\ell}(\mathbf{k};\tau_1,\tau_2)
\theta(\tau_2-\tau_1).
\label{SK_++KK}
\end{align}

%%%%%%%%%%%%%%%%%%%%%%%%%%%%%%%%%%%%%%%%%%%%%%%%%%%%%
\subsection{Inflaton Correlators from Continuum KK Gravitons}

In our setup, the inflaton \(\phi\) is localized on the brane at \(z=0\).
Its brane action is given by
\begin{align}
S_{\rm brane}
=
\int \mathrm{d}^5x\,\delta(z)\sqrt{-g_{\rm ind}}
\left[
-\frac12 g_{\rm ind}^{\mu\nu}
\partial_\mu\phi\partial_\nu\phi
-
V(\phi)
\right],
\end{align}
where \(g_{\rm ind}\) is the induced metric on the brane.
Expanding the metric in Eq.~\eqref{h_fluctuation} and writing
\(\phi=\phi_0(t)+\varphi\), we obtain the interaction between
the inflaton fluctuation and the tensor perturbation,
\begin{align}
S_{\rm int}
=
\frac12
\int \mathrm{d}\tau\,\mathrm{d}^3x\,\mathrm{d}z\,
\delta(z)\,
e^{-A(0)}
h_{ij}(\tau,\mathbf{x},0)
\partial_i\varphi
\partial_j\varphi .
\end{align}
Here we have used the transverse-traceless condition~\eqref{TT}.  We also note
that \(A(0)=0\) at the brane.  Applying the KK decomposition~\eqref{KK_0},
and using \(h_{ij}(\tau,\mathbf{x},0)=\tilde h_{ij}(\tau,\mathbf{x},0)\) at the
brane, we obtain
\begin{align}
S_{\rm int}
=
\frac12
\int \mathrm{d}\tau\,\mathrm{d}^3x
\left[
\psi_0(0)\hat h^{(0)}_{ij}
+
\int_{3H/2}^\infty \mathrm{d}m\,\psi_m(0)\hat h^{(m)}_{ij}
\right]
\partial_i\varphi
\partial_j\varphi .
\label{int_KK_phi}
\end{align}
Therefore, the effective coupling of each KK graviton to the brane-localized
inflaton is controlled by the value of its wavefunction at the brane,
\(\psi_m(0)\).  In the exchange diagram, this dependence appears through the
KK spectral weight \(\rho_{\rm KK}(m^2)\propto |\psi_m(0)|^2\).

The interaction~\eqref{int_KK_phi} contributes to the inflaton four-point
correlator through massless and massive graviton exchange.  The massless
graviton exchange contribution was computed in Ref.~\cite{Seery:2008ax}, while
massive spin-2 exchange in de Sitter spacetime has been discussed in
Refs.~\cite{Tong:2022cdz,Kumar:2018jxz,Kumar:2025anx}.  Here we focus on the contribution
from the continuum KK gravitons.

We denote the \(s\)-channel exchanged momentum by
\begin{align}
\mathbf{s}\equiv \mathbf{k}_1+\mathbf{k}_2,
\qquad
s=|\mathbf{s}|.
\end{align}
The KK graviton exchange contribution is given by
\begin{align}
\nonumber
\left\langle
\varphi_{\mathbf{k}_1}
\varphi_{\mathbf{k}_2}
\varphi_{\mathbf{k}_3}
\varphi_{\mathbf{k}_4}(0)
\right\rangle^{\prime}_{\rm KK}
=\;&
\kappa_5^2
\int_{-\infty}^0\mathrm{d}\tau_1\int_{-\infty}^0\mathrm{d}\tau_2
\sum_{\mathrm{a},\mathrm{b}=\pm}
(-\mathrm{a}\mathrm{b})
\\
\nonumber
&\times
K_{\mathrm{a}}(k_1;\tau_1)
K_{\mathrm{a}}(k_2;\tau_1)
K_{\mathrm{b}}(k_3;\tau_2)
K_{\mathrm{b}}(k_4;\tau_2)
\\
&\times
\Pi_{ij,k\ell}(\hat{\mathbf{s}})
D_{\mathrm{a}\mathrm{b}}^{\rm KK}
(s;\tau_1,\tau_2)
\mathbf{k}_{1 i} \mathbf{k}_{2 j} \mathbf{k}_{3 k} \mathbf{k}_{4 \ell}
+t+u .
\label{4pt_KK_inin}
\end{align}
Here the factor \(\kappa_5^2\) restores the overall gravitational
normalization, while \(D_{\mathrm{a}\mathrm{b}}^{\rm KK}\) denotes the reduced
brane-to-brane SK propagator for the continuum KK gravitons defined in
Eq.~\eqref{D_KK_minusplus}.  The inflaton bulk-to-boundary propagator
\(K_{\mathrm a}\) is defined in Eq.~\eqref{BB}.

For the KK continuum, the reduced brane-to-brane propagator
\(D^{\mathrm{a}\mathrm{b}}_{\rm KK}\) is dimensionless.  We therefore define
dimensionless KK seed integrals by
\begin{align}
\nonumber\mathcal I_{\mathrm{a}\mathrm{b},{\rm KK}}^{p_1p_2}
\equiv
&\ \frac{s^{5+p_1+p_2}}{H^3}
(-\mathrm{a}\mathrm{b})\\
&\times
\int_{-\infty}^{0}\mathrm{d}\tau_1\int_{-\infty}^0\mathrm{d}\tau_2\,
(-\tau_1)^{p_1}
(-\tau_2)^{p_2}
e^{i\mathrm{a}k_{12}\tau_1+i\mathrm{b}k_{34}\tau_2}
D^{\mathrm{a}\mathrm{b}}_{\rm KK}(s;\tau_1,\tau_2).
\label{def_seed_KK}
\end{align}
This definition differs from the scalar seed definition in
Eq.~\eqref{def_seed} by one extra factor of \(1/H\), reflecting the different
mass dimension of the reduced KK propagator.\footnote{The time dependence of the massive helicity-\(\pm2\) mode differs from that of
the scalar mode by two powers of conformal time at each vertex.  As a result,
the KK seed can be obtained from the scalar seed as
\begin{align}
\mathcal I_{\mathrm{a}\mathrm{b},{\rm KK}}^{p_1p_2}
=
\frac{s^4}{H^4}
\mathcal I_{\mathrm{a}\mathrm{b},{\rm scalar}}^{p_1-2,p_2-2}
\left[
\rho_{\rm scalar}
\rightarrow
\frac{\rho_{\rm KK}}{H}
\right].
\label{seed_KK_scalar_relation}
\end{align}} 
Based on this, the correlator can be written as
\begin{align}
\nonumber
\left\langle
\varphi_{\mathbf{k}_1}
\varphi_{\mathbf{k}_2}
\varphi_{\mathbf{k}_3}
\varphi_{\mathbf{k}_4}(0)
\right\rangle^{\prime}_{\rm KK}
=
&\ 
\frac{\kappa_5^2H^{11}}{16 k_1^3 k_2^3 k_3^3 k_4^3} \frac{ \mathbf{k}_{1 i} \mathbf{k}_{2 j} \mathbf{k}_{3 k} \mathbf{k}_{4 \ell}}{s^5}
\ \Pi_{ij,k\ell}(\hat{\mathbf{s}})
\\
\nonumber
&\times
\sum_{\mathrm{a},\mathrm{b}=\pm}
\biggl[
\mathcal{I}_{\mathrm{a}\mathrm{b},{\rm KK}}^{00}
+
i\mathrm{a}\,\frac{k_{12}}{s}\mathcal{I}_{\mathrm{a}\mathrm{b},{\rm KK}}^{10}
+
i\mathrm{b}\,\frac{k_{34}}{s}\mathcal{I}_{\mathrm{a}\mathrm{b},{\rm KK}}^{01}
-
\frac{k_1k_2}{s^2}\mathcal{I}_{\mathrm{a}\mathrm{b},{\rm KK}}^{20}
\\
\nonumber
&\hspace{2.0cm}
-
\frac{k_3k_4}{s^2}\mathcal{I}_{\mathrm{a}\mathrm{b},{\rm KK}}^{02}-
\mathrm{a}\mathrm{b}\,
\frac{k_{12}k_{34}}{s^2}
\mathcal{I}_{\mathrm{a}\mathrm{b},{\rm KK}}^{11}
-
i\mathrm{b}\,
\frac{k_1k_2k_{34}}{s^3}
\mathcal{I}_{\mathrm{a}\mathrm{b},{\rm KK}}^{21}
\\
&\hspace{2.0cm}
-
i\mathrm{a}\,
\frac{k_{12}k_3k_4}{s^3}
\mathcal{I}_{\mathrm{a}\mathrm{b},{\rm KK}}^{12}+
\frac{k_1k_2k_3k_4}{s^4}
\mathcal{I}_{\mathrm{a}\mathrm{b},{\rm KK}}^{22}
\biggr]
+t+u .
\label{4pt_KK}
\end{align}
Compared to the scalar example discussed around Eq.~\eqref{4pt} with the simple
interaction~\eqref{3vertex}, the tensor exchange contains several seed
integrals, together with angular dependence from the helicity-\(\pm2\)
polarization sum.

To highlight the effect of the KK spectral weight, we focus on the seed
function rather than the full four-point correlator.  The full correlator in
Eq.~\eqref{4pt_KK} contains the helicity-\(\pm2\) polarization sum and
additional momentum factors from the derivative vertices.  However, the
continuum-smearing mechanism is already captured by the spectral integral
appearing in the seed functions.  The full momentum dependence can be
reconstructed from Eq.~\eqref{4pt_KK} once the relevant seed integrals are
specified.

We decompose the KK seed into signal and background parts as in
Eq.~\eqref{Seed_dec}.  In analogy with Eq.~\eqref{def_NF}, we define the
rescaled signal by
\begin{align}
\mathcal{F}_{\rm Signal}^{\rm KK}
\equiv
\frac{
\mathcal{I}_{\rm Signal}^{\rm KK}
}{
4\pi \left(s/H\right)^4
r_1^{\frac{1}{2}+p_1}
r_2^{\frac{1}{2}+p_2}
}.
\label{def_NF_KK}
\end{align}
The overall sign is chosen for plotting convenience.
Although the Fourier--Laplace expression in Eq.~\eqref{F_signal} was obtained
using the soft-limit and large-\(\mu\) approximations, the result shown below
does not rely on these approximations.  We numerically integrate the exact
fixed-mass signal in Eqs.~\eqref{I_signal}--\eqref{def_F^p_nu} against the
exact KK spectral weight obtained from the five-dimensional wavefunctions.
Thus the smooth, non-oscillatory behavior can also be seen in the exact
spectral integral.

\begin{figure}[H]
  \begin{center}
   \includegraphics[width=100mm]{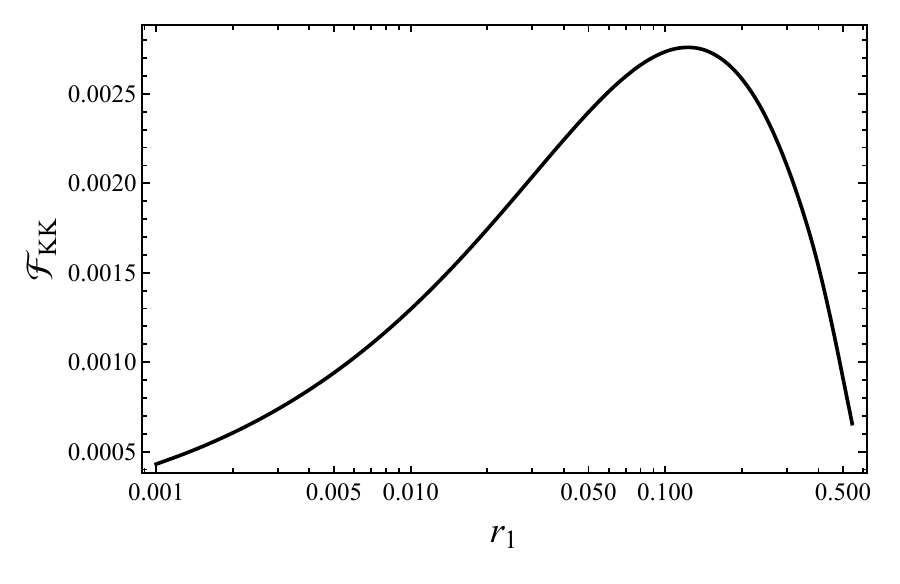}
  \end{center}
\caption{
KK continuum seed function as a function of \(r_1\).
We fix \(r_2=2/3\), \(p_1=p_2=0\), and \(\alpha=0.01\).
We plot \(\mathcal F_{\rm Signal}^{\rm KK}\): the rescaled signal part defined in
Eq.~\eqref{def_NF_KK}. The smooth,
non-oscillatory behavior reflects the smearing of the logarithmic clock signal
by the continuum KK spectrum.
}
\label{seedKK}
\end{figure}

In Fig.~\ref{seedKK}, we show the rescaled signal part
\(\mathcal F_{\rm Signal}^{\rm KK}\) defined above, for
\((p_1,p_2)=(0,0)\) as an example.  The resulting seed function is smooth as a
function of \(r_1\), without a persistent logarithmic oscillation.  This reflects
the continuum smearing of the spin-2 clock signal.  Each massive KK mode carries
a clock phase controlled by its mass, or equivalently by \(\mu\), but the
spectral integral superposes modes with different frequencies and dephases the
individual oscillations.

This behavior is also consistent with the threshold structure of the RS2-like
continuum.  The continuum starts at the principal-series edge, \(\mu=0\), so
there is no nonzero endpoint frequency that could survive as a residual
logarithmic oscillation.  Moreover, the KK spectral weight vanishes near the
threshold as \(\rho_{\rm KK}/H\propto\mu\), which suppresses the deep-soft
contribution.  The KK continuum therefore produces a smooth broad feature
rather than a sharp fixed-frequency cosmological-collider clock.

Let us finally comment on analytic background contributions.  In
Fig.~\ref{seedKK}, we have shown only the non-analytic signal part of the KK
seed function.  As discussed in Sec.~\ref{sec.BG}, analytic background terms
are local from the four-dimensional low-energy viewpoint and do not constitute
the characteristic cosmological-collider clock signal.  Although their
coefficients are fixed in principle by the five-dimensional geometry in the
RS2-like realization, we separate them from the universal non-analytic signal
and do not include them in the figure.

% ============================================================
\section{Relation to Conformal Stress-Tensor Continua}
\label{sec:CFT_relation}
% ============================================================

The smooth KK signal found in the previous section motivates a comparison
with conformal spectator sectors.\footnote{We thank the anonymous referee for pointing out this useful connection.}  A useful framework for this purpose is
the de Sitter K\"all\'en--Lehmann representation of
Ref.~\cite{Loparco:2023rug}, in which the two-point function of a local
operator is decomposed into propagators associated with unitary
representations of the de Sitter isometry group.

For a generic symmetric traceless rank-two operator, the decomposition
contains principal-series representations with spins
\(\ell=0,1,2\).  For the conserved stress tensor of a four-dimensional CFT,
\begin{equation}
    \Delta_T=4,
    \qquad
    \nabla^\mu T_{\mu\nu}=0,
\end{equation}
the spin-zero and spin-one spectral densities vanish, leaving only the spin-two principal
series. Here \(\ell\) labels the spin of the de Sitter representation; each massive spin-two
representation still contains helicities \(0,\pm1,\pm2\).
Its two-point function can therefore be written schematically as
\begin{equation}
    \langle
    T_{\mu\nu}(Y_1)T_{\rho\sigma}(Y_2)
    \rangle
    =
    \int_0^\infty d\mu\,
    \rho_T^{(2)}(\mu)\,
    G_{\mu,2;\mu\nu,\rho\sigma}(Y_1,Y_2),
    \label{eq:CFT_KL}
\end{equation}
where \(G_{\mu,2}\) is the propagator associated with a principal-series
spin-two representation.  For the spin-two principal series in four-dimensional de Sitter space,
\begin{equation}
    \frac{m^2}{H^2}
    =
    \mu^2+\frac94 ,
    \qquad
    \mu\in\mathbb{R}_{\ge0}.
    \label{eq:CFT_mass}
\end{equation}
Since the K\"all\'en--Lehmann representation in
Eq.~\eqref{eq:CFT_KL} integrates over \(\mu\in[0,\infty)\), the
principal-series continuum starts at \(m=3H/2\), corresponding to
zero clock frequency,~\(\mu=0\).

For this conserved CFT stress tensor, the corresponding spectral density is
\cite{Loparco:2023rug}
\begin{equation}
    \rho_T^{(2)}(\mu)
    =
    c_T\,\frac{\pi}{320}
    \left|
        \Gamma\left(\frac52+i\mu\right)
    \right|^2
    \mu\sinh(\pi\mu).
    \label{eq:CFT_rho}
\end{equation}
Near the lower endpoint,
\begin{equation}
    \rho_T^{(2)}(\mu)
    =
    c_T\,\frac{9\pi^3}{5120}\,
    \mu^2
    +\mathcal O(\mu^4).
    \label{eq:CFT_rho_IR}
\end{equation}

This can be compared directly with the RS2 spectral representation.
Since $dm^2/H^2
    =
    2\mu\,d\mu $,
the KK weight with respect to the \(d\mu\) measure is
\begin{equation}
    W_{\rm KK}(\mu)
    \equiv
    2\mu\,\frac{\rho_{\rm KK}(\mu)}{H}.
    \label{eq:WKK}
\end{equation}
Using the threshold behavior in Eq.~\eqref{rho_KK_asmp}, we find
\begin{equation}
    W_{\rm KK}(\mu)\propto\mu^2,
    \qquad
    \mu\to0.
    \label{eq:WKK_IR}
\end{equation}
Thus the CFT stress-tensor continuum and the RS2 KK continuum share both
the same lower endpoint, \(\mu_0=0\), and the same quadratic onset of the
effective spectral weight.

Away from the endpoint, however, their spectral distributions differ.
In particular,
\begin{equation}
    \rho_T^{(2)}(\mu)\propto\mu^5
    \qquad (\mu\gg1),
\end{equation}
whereas
\begin{equation}
    W_{\rm KK}(\mu)\longrightarrow\frac{2}{\pi}
\end{equation}
in the large-mass regime \(m/K\gg1\).
This common threshold behavior therefore does not distinguish the two
continua, while the full spectral distribution remains sensitive to the
microscopic realization of the continuum.

For a schematic coupling
\(\nabla_\mu\phi\nabla_\nu\phi\,T^{\mu\nu}\), the
K\"all\'en--Lehmann representation expresses the corresponding contribution
as a spectral integral over fixed-mass spin-two exchange.
To compare the resulting non-analytic signals, we restrict to the
helicity-\(\pm2\) sector and use the same fixed-mass seed as in the KK
calculation. With the normalization of Eq.~\eqref{def_NF_KK}, we define
\begin{equation}
    \mathcal F_{\rm Signal}^{\rm CFT}
    \equiv
    \frac{\mathcal I_{\rm Signal}^{\rm CFT}}
    {4\pi (s/H)^4
    r_1^{1/2+p_1}r_2^{1/2+p_2}} .
    \label{def_NF_CFT}
\end{equation}
In analogy with Eq.~\eqref{seed_KK_scalar_relation},
\begin{equation}
    \mathcal I_{\rm Signal}^{\rm CFT}
    \equiv
    \frac{s^4}{H^4}
    \mathcal I_{\rm Signal,scalar}^{p_1-2,p_2-2}
    \left[
        \rho_{\rm scalar}
        \rightarrow
        \frac{\rho_T^{(2)}(\mu)}{2\mu}
    \right].
    \label{I_CFT_signal}
\end{equation}
The factor \(1/(2\mu)\) simply converts the
\(dm^2/H^2=2\mu\,d\mu\) measure used in our seed representation
to the \(d\mu\) measure of the CFT K\"all\'en--Lehmann representation.
Up to a \(\mu\)-independent polarization normalization, the
helicity-\(\pm2\) projection of \(G_{\mu,2}\) uses the same
Bunch--Davies normalized massive spin-two mode as in the KK calculation.

\begin{figure}[t]
    \centering

    \begin{subfigure}[t]{0.48\textwidth}
        \centering
        \includegraphics[width=\textwidth]{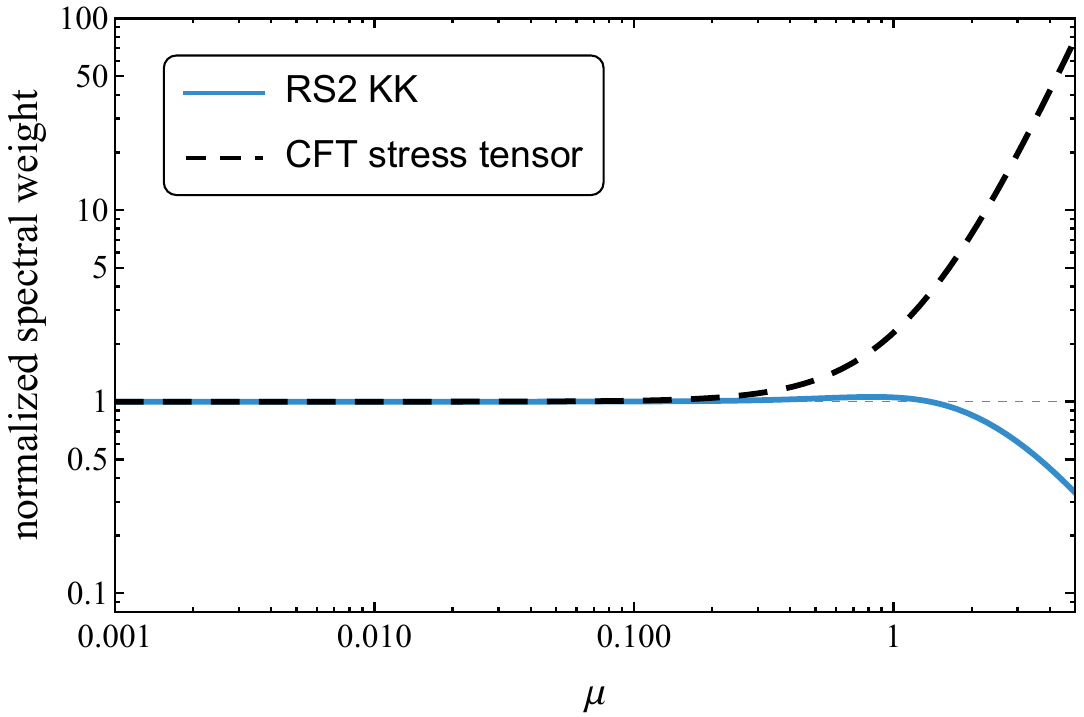}
        \caption{Normalized spectral weights.}
        \label{fig:CFT_KK_weight}
    \end{subfigure}
    \hfill
    \begin{subfigure}[t]{0.48\textwidth}
        \centering
        \includegraphics[width=\textwidth]{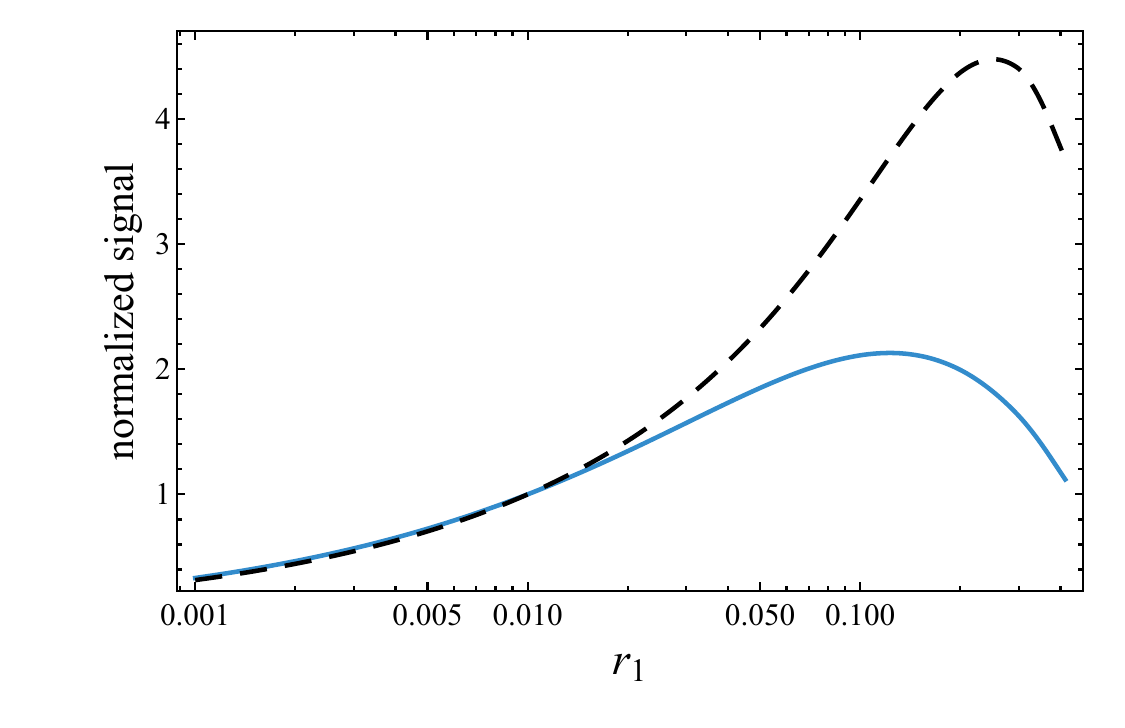}
        \caption{Normalized helicity-\(\pm2\) signals.}
        \label{fig:CFT_KK_seed}
    \end{subfigure}

  \caption{
Comparison between the RS2 KK continuum and the conserved-CFT
stress-tensor continuum.
Left: the effective spectral weights~\eqref{eq:WKK} and
\eqref{eq:CFT_rho}, each divided by
its full leading small-\(\mu\) asymptotic expression, so that both
approach unity as \(\mu\to0\).
Right: the corresponding non-analytic helicity-\(\pm2\) signals~\eqref{def_NF_KK} and~\eqref{def_NF_CFT},
obtained from the same fixed-mass seed with the two different spectral
weights.
We set \(p_1=p_2=0\) and \(r_2=2/3\), take
\(\alpha=0.01\) for the RS2 result, and normalize the two signals at
\(r_1=0.01\).
}
    \label{fig:CFT_KK_comparison}
\end{figure}

As shown in Fig.~\ref{fig:CFT_KK_comparison}, the two spectral weights
have the same threshold scaling but differ substantially away from
\(\mu=0\).  In the deep-soft limit, the spectral integral is increasingly
dominated by the vicinity of the endpoint \(\mu=0\), as contributions
from finite \(\mu\) increasingly cancel due to dephasing. Since the two
effective spectral weights have the same quadratic onset and multiply
the same fixed-mass helicity-\(\pm2\) seed, the resulting non-analytic
signals exhibit the same qualitative deep-soft behavior, while their
finite-momentum shapes separate. The absence of a persistent logarithmic
clock in the RS2 model can therefore be understood in terms of a more
general endpoint structure, also realized by the conformal stress-tensor
continuum: the continuum starts at zero clock frequency and its effective
spectral weight vanishes at the endpoint. The finite-momentum dependence,
on the other hand, retains information about the microscopic realization
of the continuum.

Inflationary correlators mediated by a conserved conformal stress tensor
have also been studied explicitly in Ref.~\cite{Pimentel:2025rds}.
A comparison with the complete stress-tensor exchange would additionally
require the helicity-\(\pm1\) and helicity-0 components, whereas the
present analysis is restricted to the helicity-\(\pm2\) sector.

%%%%%%%%%%%%%%%%%%%%%%%%%%%%%%%%%%%%%%%%%%%%%%%%%
\section{Conclusion}
\label{sec:Conclusion}

In this work, we studied how a continuous mass spectrum modifies the standard
cosmological-collider signal, generalizing the previous work~\cite{Aoki:2023tjm}.  The usual signal from a single massive particle
contains a logarithmic clock oscillation whose frequency is fixed by the mass of
the exchanged particle.  We emphasized that, once the exchanged state is
replaced by a continuum, the corresponding correlator is obtained by averaging
over a continuous set of clock frequencies.  This spectral average leads to
dephasing of the individual oscillatory components and can smear the sharp
single-particle clock signal.

We first developed this idea in a scalar toy model.  We introduced a massive
scalar field with a spectral function \(\rho(m^2)\) and studied the inflaton
four-point function mediated by its exchange.  The non-analytic part of the seed
function can be written as a mass integral over the usual fixed-mass seed.  In
the soft limit, this integral takes the form of a Fourier--Laplace transform of
the effective spectral weight with respect to logarithmic momentum variables.
This representation makes the smearing mechanism transparent.  For a continuum
with a nonzero principal-series threshold, an endpoint oscillation with the
threshold frequency may remain, although its amplitude is damped.  On the other
hand, if the continuum starts at zero clock frequency, the endpoint oscillation
itself disappears, and the signal becomes a smooth broad feature rather than a
persistent logarithmic clock.

We then realized this mechanism in an RS2-like inflationary braneworld.  In this
setup, our four-dimensional inflating universe is identified with a de Sitter
brane embedded in a five-dimensional AdS bulk.  The tensor sector contains a
localized graviton zero mode and a continuum of KK gravitons.  From the
four-dimensional point of view, the KK tower is a continuum of massive spin-2
states starting at the principal-series threshold \(m=3H/2\), or equivalently
\(\mu=0\).  We derived the corresponding KK wavefunctions and identified the
effective brane spectral weight,
\[
\rho_{\rm KK}(m^2)
=
\frac{H^2}{2m}
|\psi_m(0)|^2 .
\]
Unlike the phenomenological spectral functions considered in the toy model,
\(\rho_{\rm KK}\) is fixed by the five-dimensional geometry and by the value of
the KK wavefunction on the brane.  We found that the dimensionless spectral
weight behaves as \(\rho_{\rm KK}/H\propto\mu\) near the threshold, vanishes at
\(\mu=0\), and decreases in the large-mass regime.

Finally, we applied the KK spectral representation to the inflaton four-point
function.  Since the inflaton is localized on the brane, the coupling of each KK
graviton to inflaton fluctuations is controlled by the same brane wavefunction
factor \(\psi_m(0)\).  The resulting exchange diagram is therefore described by
a brane-to-brane propagator with the KK spectral weight.  We showed that the
corresponding seed function is smooth in the soft region and does not exhibit a
persistent logarithmic oscillation.  This is the RS2 realization of the
continuum-smearing mechanism: the continuum KK gravitons superpose spin-2 clock
signals with different masses, and the integration over the continuum dephases
the individual oscillations. We also compared the RS2 KK continuum with a conserved conformal
stress-tensor continuum in the helicity-\(\pm2\) sector.  The two share
the same threshold at \(\mu=0\) and the same quadratic onset of the
effective spectral weight, while their finite-momentum shapes differ.

The analysis in this paper illustrates that deviations from the standard
cosmological-collider clock signal can carry information about the structure of
the high-energy spectrum.  In particular, the absence of a sharp oscillatory
signal does not necessarily imply the absence of heavy particles during
inflation.  It may instead indicate that the relevant degrees of freedom form a
continuum, or a sufficiently dense tower, whose different clock frequencies are
smeared in primordial correlators.  This provides a useful perspective for
probing extra dimensions and other UV sectors through cosmological correlation
functions.

Several directions remain open.  First, it would be interesting to extend the
analysis beyond the helicity-\(\pm2\) sector and include the helicity-0 and
helicity-\(\pm1\) components of the massive spin-2 KK modes.  These components
are expected to affect the full angular and soft-momentum structure of the
correlator, and their inclusion is necessary for a complete treatment of
massive spin-2 exchange.

Second, it would be important to study more realistic slow-roll backgrounds,
the conversion to curvature perturbations, and observational templates for CMB
and large-scale-structure analyses.  Even when the sharp logarithmic
oscillation is smeared or absent, the continuum can leave characteristic soft
scalings and broad momentum-space features.  Such features may provide
observational handles complementary to the usual fixed-frequency clock
templates.

Another interesting direction is the tensor sector.  Higher-dimensional
inflationary scenarios can modify tensor perturbations, and related
``transdimensional'' effects have been discussed in
Refs.~\cite{Giudice:2002vh,Im:2017eju}.  It would be worthwhile to understand
whether the continuum KK graviton spectrum studied here leaves distinctive
imprints in tensor correlators or mixed scalar-tensor correlators.

Finally, an IR-brane regularization of the continuum, or more general dense
discrete spectra, could provide a useful bridge between isolated
cosmological-collider particles and genuinely continuous spectra.  This would
also clarify how the continuum-smearing behavior emerges from a sequence of
discrete KK resonances.

\paragraph{Acknowledgments}
We thank Zhehan Qin, Yi Wang, and Yuhang Zhu for helpful discussion. S.A. is supported by the Japan Science and Technology Agency (JST) 
as part of Adopting Sustainable Partnerships for Innovative Research Ecosystem (ASPIRE), 
grant JPMJAP2318, and by JSPS KAKENHI Grant Number 26K07078.

\begin{appendix}

\section{Calculation of the Seed Function}
\label{MBdetail}

Here we present some details of the computation of the seed
functions~\eqref{def_seed}.  The main obstacle is to perform the nested time
integrals involving the product of Hankel functions contained in the massive
propagator \(D_{\mathrm{ab}}\).  To evaluate these integrals, we use the
Mellin--Barnes (MB) representation of the Hankel function,
\begin{align}
H_\nu^{(j)}(-k\tau)
=
\int_{-i\infty}^{i\infty}
\frac{\mathrm{d}z}{2\pi i}\,
\frac{(-k\tau/2)^{-2z}}{\pi}
e^{(-1)^{j+1} i\frac{\pi}{2}(2z-\nu-1)}
\Gamma\left(z-\frac{\nu}{2}\right)
\Gamma\left(z+\frac{\nu}{2}\right),
\qquad
(j=1,2).
\end{align}
This representation allows us to perform the time integrals explicitly.  The
remaining complex \(z\)-integrals can then be evaluated by closing the contours
and summing over the corresponding residues.

\subsubsection*{Wightman Contributions}
Let us start with the Wightman propagators,
\((\mathrm{a},\mathrm{b})=(\pm,\mp)\).  Using the MB representation above, the
propagator \(D_{\pm\mp}\) can be written as
\begin{align}
\nonumber
D_{\pm\mp}(s;\tau_1,\tau_2)
=&\;
\frac{H^2}{4\pi}
\int_{m^2}
\int_{-i\infty}^{i\infty}
\frac{\mathrm{d}z_1}{2\pi i}
\frac{\mathrm{d}z_2}{2\pi i}
\left(\frac{s}{2}\right)^{-2(z_1+z_2)}
\left(-\tau_1\right)^{-2z_1+\frac32}
\left(-\tau_2\right)^{-2z_2+\frac32}
e^{\mp i\pi(z_1-z_2)}
\\
&\times
\Gamma\left(z_1-\frac{\nu}{2}\right)
\Gamma\left(z_1+\frac{\nu}{2}\right)
\Gamma\left(z_2-\frac{\nu}{2}\right)
\Gamma\left(z_2+\frac{\nu}{2}\right).
\label{D_W_MB}
\end{align}

Substituting Eq.~\eqref{D_W_MB} into
the seed,
\begin{align}
\mathcal I_{\pm\mp}^{p_1p_2}
=
&\,
\frac{s^{5+p_1+p_2}}{H^2}
\int_{-\infty}^{0}\mathrm{d}\tau_1
\int_{-\infty}^{0}\mathrm{d}\tau_2\,
(-\tau_1)^{p_1}
(-\tau_2)^{p_2}
e^{\pm i k_{12}\tau_1}
e^{\mp i k_{34}\tau_2}
D_{\pm\mp}(s;\tau_1,\tau_2),    
\end{align}
the time integrals are evaluated using
\begin{align}
\int_{-\infty}^{0}\mathrm{d}\tau\,
e^{\pm i k\tau}
(-\tau)^{p-1}
=
e^{\mp i\frac{\pi}{2}p}
k^{-p}
\Gamma(p),
\end{align}
with the usual \(i\epsilon\) prescription.  We then obtain
\begin{align}
\nonumber
\mathcal I_{\pm\mp}^{p_1p_2}
=&\;
\frac{
r_1^{\frac52+p_1}
r_2^{\frac52+p_2}
}{4\pi}
e^{\mp i\frac{\pi}{2}(p_1-p_2)}
\int_{m^2}
\int_{-i\infty}^{i\infty}
\frac{\mathrm{d}z_1}{2\pi i}
\frac{\mathrm{d}z_2}{2\pi i}
\left(\frac{r_1}{2}\right)^{-2z_1}
\left(\frac{r_2}{2}\right)^{-2z_2}
\\
\nonumber&\times
\Gamma\left(z_1-\frac{\nu}{2}\right)
\Gamma\left(z_1+\frac{\nu}{2}\right)
\Gamma\left(z_2-\frac{\nu}{2}\right)
\Gamma\left(z_2+\frac{\nu}{2}\right)\\
&\times 
\Gamma\left(p_1+\frac52-2z_1\right)
\Gamma\left(p_2+\frac52-2z_2\right),
\label{I_W_MB_integral}
\end{align}
where we introduced the momentum ratio
\begin{align}
r_1 \equiv \frac{s}{k_{12}},
\qquad
r_2 \equiv \frac{s}{k_{34}}.
\end{align}
For \(0<r_1,r_2<1\), we close the \(z_1\) and \(z_2\) contours to the
left.  The poles picked up in this way are
\begin{align}
z_i=\pm\frac{\nu}{2}-n_i,
\qquad
n_i=0,1,2,\cdots .
\end{align}
The remaining poles,
\begin{align}
z_i=\frac{1}{2}\left(p_i+\frac52+n_i\right),
\qquad
n_i=0,1,2,\cdots ,
\end{align}
lie on the other side of the contour.  Thus the MB contours should be chosen
inside the corresponding fundamental strips.  In particular, for both the
principal and complementary series, a convenient choice is
\begin{align}
\frac{|\mathrm{Re}\,\nu|}{2}
<
\mathrm{Re}\,z_i
<
\frac{\mathrm{Re}\,p_i+\frac52}{2}.
\end{align}

Summing over these residues gives
\begin{align}
\mathcal I_{\pm\mp}^{p_1p_2}
=
\frac{
e^{\mp i\frac{\pi}{2}(p_1-p_2)}
}{4\pi}
\int_{m^2}
\left[
\left(\frac{r_1}{2}\right)^{\nu}
\mathbf{F}_{\nu}^{p_1}(r_1)
+
(\nu\rightarrow-\nu)
\right]
\left[
\left(\frac{r_2}{2}\right)^{\nu}
\mathbf{F}_{\nu}^{p_2}(r_2)
+
(\nu\rightarrow-\nu)
\right],
\label{I_pmmp_result_bar_nu}
\end{align}
where
\begin{align}
\mathbf{F}_{\nu}^{p}(r)
\equiv
r^{\frac{5}{2}+p}
\Gamma\left(\frac{5}{2}+p+ \nu\right)
\Gamma\left(- \nu\right)
{}_2 \mathrm{F}_1
\left[
\left.
\begin{array}{c}
\frac{5}{4}+\frac{p}{2}+\frac{\nu}{2},
\frac{7}{4}+\frac{p}{2}+\frac{\nu}{2}
\\
1+ \nu
\end{array}
\right\rvert\,
r^2
\right]. \label{def_F(r)}
\end{align}
Here, ${}_2 \mathrm{F}_1$ denotes the hypergeometric function.

%％％％％％％％％％％％％％％％％％％％％％％％％％％％％％％％％

\subsubsection*{Time-Ordered Contributions}
We next consider the time-ordered and anti-time-ordered propagators,
\((\mathrm{a},\mathrm{b})=(\pm,\pm)\).  Following Refs.~\cite{Qin:2022fbv,Qin:2023ejc}, and assuming
\(r_1<r_2\), we decompose the propagator as
\begin{align}
D_{\pm\pm}(s;\tau_1,\tau_2)
=
D_{\mp\pm}(s;\tau_1,\tau_2)
+
\left[
D_{\pm\mp}(s;\tau_1,\tau_2)
-
D_{\mp\pm}(s;\tau_1,\tau_2)
\right]
\theta(\tau_2-\tau_1).
\label{D_toto_decomp}
\end{align}
The first term is factorized in the two time variables, while the second term
contains the genuine time-ordering effect.  Accordingly, we write
\begin{align}
\mathcal I_{\pm\pm}^{p_1p_2}
=
\mathcal I_{\pm\pm,\mathrm F}^{p_1p_2}
+
\mathcal I_{\pm\pm,\mathrm{TO}}^{p_1p_2},
\end{align}
where
\begin{align}
\nonumber
\mathcal I_{\pm\pm,\mathrm F}^{p_1p_2}
\equiv
-&\,
\frac{s^{5+p_1+p_2}}{H^2}
\int_{-\infty}^{0}\mathrm{d}\tau_1
\int_{-\infty}^{0}\mathrm{d}\tau_2\,
(-\tau_1)^{p_1}
(-\tau_2)^{p_2}
e^{\pm i k_{12}\tau_1}
e^{\pm i k_{34}\tau_2}
D_{\mp\pm}(s;\tau_1,\tau_2),
\\
\nonumber
\mathcal I_{\pm\pm,\mathrm{TO}}^{p_1p_2}
\equiv
-&\,
\frac{s^{5+p_1+p_2}}{H^2}
\int_{-\infty}^{0}\mathrm{d}\tau_1
\int_{-\infty}^{0}\mathrm{d}\tau_2\,
(-\tau_1)^{p_1}
(-\tau_2)^{p_2}
e^{\pm i k_{12}\tau_1}
e^{\pm i k_{34}\tau_2}
\\
&\times
\left[
D_{\pm\mp}(s;\tau_1,\tau_2)
-
D_{\mp\pm}(s;\tau_1,\tau_2)
\right]
\theta(\tau_2-\tau_1).
\label{I_toto_F_TO_def}
\end{align}

The factorized part can be evaluated in the same way as
\(\mathcal I_{\pm\mp}^{p_1p_2}\).  We obtain
\begin{align}
\nonumber \mathcal I_{\pm\pm,\mathrm F}^{p_1p_2}
=
\frac{
\pm i\, e^{\mp i\frac{\pi}{2}(p_1+p_2)}
}{4\pi}
\int_{m^2}
&\left[
\left(\frac{r_1}{2}\right)^{\nu}
e^{\mp i\pi\nu}
\mathbf{F}_{\nu}^{p_1}(r_1)
+
(\nu\rightarrow-\nu)
\right]\\
\times & 
\left[
\left(\frac{r_2}{2}\right)^{\nu}
\mathbf{F}_{\nu}^{p_2}(r_2)
+
(\nu\rightarrow-\nu)
\right],
\label{I_pmpm_F_result}
\end{align}
where the function $\mathbf{F}_{\nu}^{p}(r)$ is defined in Eq.~\eqref{def_F(r)}.

For the time-ordered part, the time integrals are nested.  We use the formula
\begin{align}
\nonumber
&\int_{-\infty}^{0}\mathrm{d}\tau_2
\int_{-\infty}^{\tau_2}\mathrm{d}\tau_1\,
e^{\pm i(k_{12}\tau_1+k_{34}\tau_2)}
(-\tau_1)^{p-1}
(-\tau_2)^{q-1}
\\
&\hspace{2cm}
=
e^{\mp i\frac{\pi}{2}(p+q)}
k_{12}^{-p-q}
\mathfrak F
\left[
\left.
\begin{array}{c}
q,\ p+q
\\
1+q
\end{array}
\right|-\frac{k_{34}}{k_{12}}
\right],
\end{align}
where
\begin{align}
\mathfrak F
\left[
\left.
\begin{array}{c}
a,\ b
\\
c
\end{array}
\right|z
\right]
\equiv
\frac{\Gamma(a)\Gamma(b)}{\Gamma(c)}
{}_2F_1
\left[
\left.
\begin{array}{c}
a,\ b
\\
c
\end{array}
\right|z
\right].
\end{align}

Using this formula, we obtain
\begin{align}
\nonumber
\mathcal I_{\pm\pm,\mathrm{TO}}^{p_1p_2}
=&\;
\frac{\pm i}{4\pi}
r_1^{5+p_1+p_2}
e^{\mp i\frac{\pi}{2}(p_1+p_2)}
\int_{m^2}
\int_{-i\infty}^{i\infty}
\frac{\mathrm{d}z_1}{2\pi i}
\frac{\mathrm{d}z_2}{2\pi i}
\left(\frac{r_1}{2}\right)^{-2(z_1+z_2)}
\\
\nonumber
&\times
\left(
e^{\pm 2\pi i z_2}
-
e^{\pm 2\pi i z_1}
\right)
\Gamma\left(z_1-\frac{\nu}{2}\right)
\Gamma\left(z_1+\frac{\nu}{2}\right)
\Gamma\left(z_2-\frac{\nu}{2}\right)
\Gamma\left(z_2+\frac{\nu}{2}\right)
\\
&\times
\mathfrak F
\left[
\left.
\begin{array}{c}
p_2-2z_2+\frac{5}{2},\
p_1+p_2-2(z_1+z_2)+5
\\
p_2-2z_2+\frac{7}{2}
\end{array}
\right|
-\frac{r_1}{r_2}
\right].
\label{I_toto_TO_MB}
\end{align}

The factor
\(\left(e^{\pm2\pi i z_2}-e^{\pm2\pi i z_1}\right)\) removes the contributions
from the pole families with the same sign of \(\nu\).  Therefore only the mixed
pole families contribute. Taking the residues at
\begin{align}
z_1=\frac{\nu}{2}-n_1,
\qquad
z_2=-\frac{\nu}{2}-n_2,
\qquad
n_1,n_2=0,1,2,\cdots ,
\end{align}
together with the contribution obtained by \(\nu\rightarrow-\nu\), we find
\begin{align}
\nonumber
\mathcal I_{\pm\pm,\mathrm{TO}}^{p_1p_2}
=&\;
\frac{
e^{\mp i\frac{\pi}{2}(p_1+p_2)}
}{2}
r_1^{5+p_1+p_2}
\sum_{n_1,n_2=0}^{\infty}
\frac{(-1)^{n_1+n_2}}{n_1!n_2!}
\left(\frac{r_1}{2}\right)^{2(n_1+n_2)}
\int_{m^2}
\frac{(\nu)_{-n_1}(-\nu)_{-n_2}}{-\nu}
\\
&\times
\mathfrak F
\left[
\left.
\begin{array}{c}
p_2+\frac{5}{2}+2n_2+\nu,\
p_1+p_2+5+2(n_1+n_2)
\\
p_2+\frac{7}{2}+2n_2+\nu
\end{array}
\right|
-\frac{r_1}{r_2}
\right]
+
(\nu\rightarrow-\nu).
\label{I_toto_TO_result}
\end{align}
Here \((a)_n\) denotes the Pochhammer symbol, analytically continued to
negative integer \(n\) by
\begin{align}
(a)_{-n}
=
\frac{\Gamma(a-n)}{\Gamma(a)}.
\end{align}

\subsubsection*{Signal--Background Decomposition}
Finally, we combine the Wightman pieces
\(\mathcal I_{\pm\mp}^{p_1p_2}\), the factorized pieces
\(\mathcal I_{\pm\pm,\mathrm F}^{p_1p_2}\), and the time-ordered pieces
\(\mathcal I_{\pm\pm,\mathrm{TO}}^{p_1p_2}\).  The total seed integral is
defined by
\begin{align}
\mathcal I_{\rm Total}
\equiv
\sum_{\mathrm a,\mathrm b=\pm}
\mathcal I_{\mathrm a\mathrm b}^{p_1p_2}.
\end{align}
It is useful to decompose it into the non-analytic signal part and the analytic
background part,
\begin{align}
\mathcal I_{\rm Total}
=
\mathcal I_{\rm Signal}
+
\mathcal I_{\rm BG}.
\end{align}

The signal part comes from the factorized terms and is given by
\begin{align}
\mathcal I_{\rm Signal}
=
\int_{m^2}
\mathcal C_{\nu}^{p_1p_2}
\Bigg[
&
\left(\frac{r_1r_2}{4}\right)^{\nu}
\mathbf{F}_{\nu}^{p_1}(r_1)
\mathbf{F}_{\nu}^{p_2}(r_2)
+
\left(\frac{r_1}{r_2}\right)^{\nu}
\mathbf{F}_{\nu}^{p_1}(r_1)
\mathbf{F}_{-\nu}^{p_2}(r_2)
+
(\nu\rightarrow-\nu)\Bigg]
,
\end{align}
where
\begin{align}
\mathcal C_{\nu}^{p_1p_2}
\equiv
\frac{1}{2\pi}
\left[
\cos\left(\frac{\pi}{2}(p_1-p_2)\right)
+
\sin\left(\frac{\pi}{2}(p_1+p_2+2\nu)\right)
\right].
\end{align}

The remaining contribution is the background part.  For \(r_1<r_2\), it is
written as
\begin{align}
\nonumber
\mathcal I_{\rm BG}
=&\;
\cos\left(\frac{\pi}{2}(p_1+p_2)\right)
r_1^{5+p_1+p_2}
\sum_{n_1,n_2=0}^{\infty}
\frac{(-1)^{n_1+n_2}}{n_1!n_2!}
\left(\frac{r_1}{2}\right)^{2(n_1+n_2)}
\\
\nonumber
&\times
\int_{m^2}\biggl\{
\frac{(\nu)_{-n_1}(-\nu)_{-n_2}}{-\nu}
\,
\mathfrak F
\left[
\left.
\begin{array}{c}
p_2+\frac52+2n_2+\nu,\,
p_1+p_2+5+2(n_1+n_2)
\\
p_2+\frac72+2n_2+\nu
\end{array}
\right|
-\frac{r_1}{r_2}
\right]
\\
&+
(\nu\rightarrow-\nu)\biggr\}.
\label{I_BG}
\end{align}
The signal part contains the non-analytic powers \(r_i^{\pm\nu}\), which are
responsible for the cosmological-collider signal.  By contrast,
\(\mathcal I_{\rm BG}\) is analytic in \(r_1\) for fixed \(r_1/r_2\), and is
therefore identified with the background contribution.  For \(r_2<r_1\), the
corresponding expression is obtained by exchanging
\((r_1,p_1)\leftrightarrow(r_2,p_2)\).

%%%%%%%%%%%%%%%%%%%%%%%%%%%%%%%%%%%%%%%%%%%%%%%%%%%%%%%%%%%%%%%%%%%%%%%%%%%
\section{Graviton Equation of Motion in the RS2 Model}
\label{G_EOM}

In this appendix, we summarize the derivation of the tensor-mode equation used
in Sec.~\ref{UV}.  We keep the discussion as general as possible at first:
the warp factor \(A(z)\) and the four-dimensional scale factor \(a(t)\) are not
specified until the end.  The derivation closely follows the discussion of
graviton fluctuations in warped backgrounds without cosmic expansion in
Ref.~\cite{Csaki:2000fc}; see also
Refs.~\cite{Kumar:2018jxz,Kumar:2025anx} for related discussions with cosmic
expansion.

We start from the five-dimensional Einstein equation~\eqref{Eeq}, or
\begin{align}
G_{MN}=\mathcal{T}_{MN}\label{EOM_GT}
\end{align}
where $\mathcal{T}_{MN}\equiv -\Lambda_5 g_{M N}+\kappa_5^2 T_{M N}$,
and linearize
it around a warped background.  It is convenient to factor out the warp factor
as
\begin{align}
g_{MN}=e^{-A(z)}\hat g_{MN}.
\label{Rescaling}
\end{align}
Under this conformal rescaling, the Einstein tensor is related to the hatted
one by
\begin{align}
G_{MN}
=
\hat G_{MN}
+\frac{3}{2}
\left[
\frac12 \hat\nabla_M A \hat\nabla_N A
+\hat\nabla_M\hat\nabla_N A
-\hat g_{MN}\hat g^{PQ}
\left(
\hat\nabla_P\hat\nabla_Q A
-\frac12 \hat\nabla_P A\hat\nabla_Q A
\right)
\right],
\label{E_tensor}
\end{align}
where hatted quantities are constructed from \(\hat g_{MN}\).
This formula is purely kinematical and does not rely on the explicit form of
\(A(z)\).

We then decompose the hatted metric into a background and a fluctuation,
\begin{align}
\hat g_{MN}=\bar g_{MN}+h_{MN}.
\end{align}
For the moment, \(\bar g_{MN}\) is kept general, except that it is taken to be a
direct product of a four-dimensional cosmological background and the extra
dimension,
\begin{align}
\mathrm{d}\bar s^2
=
\bar g_{\mu\nu}(x)\mathrm{d}x^\mu \mathrm{d}x^\nu
+
\mathrm{d}z^2 .
\end{align}
The inflationary RS2 background used in the main text corresponds to the
special case
\begin{align}
\bar g_{MN}
=
{\rm diag}\left(-1,a^2(t),a^2(t),a^2(t),1\right),
\label{bar_metric_FRW}
\end{align}
together with the warp factor \(A=A(z)\) given in Eq.~\eqref{sol_A}.

In principle, scalar and vector perturbations, as well as fluctuations of the
warp factor, can also be included.  In this appendix we restrict ourselves to
the pure tensor sector relevant for the main text.  Thus we take the only
non-vanishing components of the metric fluctuation to be \(h_{ij}\),
\begin{align}
h_{00}=h_{0i}=h_{0z}=h_{iz}=h_{zz}=0,
\qquad
h_{ij}\neq0,\label{h_ij_2}
\end{align}
and impose the transverse-traceless conditions on the four-dimensional spatial
indices,
\begin{align}
\partial^i h_{ij}=0,
\qquad
\delta^{ij}h_{ij}=0.\label{TT_2}
\end{align}
Equivalently, we are projecting the full perturbation equations onto the
helicity-\(\pm2\) sector.

We now expand Eq.~\eqref{E_tensor} to first order in \(h_{MN}\).
The following standard identities will be used.  The Christoffel symbol of
\(\hat g_{MN}\) is defined by
\begin{align}
\hat{\Gamma}_{MN}^{P}
\equiv
\frac{1}{2}\hat g^{PQ}
\left(
\partial_M \hat g_{NQ}
+\partial_N \hat g_{MQ}
-\partial_Q \hat g_{MN}
\right).
\end{align}
Its linear perturbation is
\begin{align}
\delta \hat{\Gamma}_{MN}^{P}
\equiv
\hat{\Gamma}_{MN}^{P}
-\bar{\Gamma}_{MN}^{P}
=
\frac{1}{2}
\left(
\bar{\nabla}_M h_N{}^{P}
+\bar{\nabla}_N h_M{}^{P}
-\bar{\nabla}^{P} h_{MN}
\right),
\end{align}
where indices are raised and lowered by the background metric
\(\bar g_{MN}\), and \(\bar\nabla\) denotes the corresponding covariant
derivative.  Similarly, for the Riemann tensor,
\begin{align}
\hat{R}^{R}{}_{SMN}
\equiv
\partial_M \hat{\Gamma}_{NS}^{R}
-\partial_N \hat{\Gamma}_{MS}^{R}
+\hat{\Gamma}_{MQ}^{R}\hat{\Gamma}_{NS}^{Q}
-\hat{\Gamma}_{NQ}^{R}\hat{\Gamma}_{MS}^{Q},
\end{align}
we have
\begin{align}
\delta \hat{R}^{R}{}_{SMN}
\equiv
\hat{R}^{R}{}_{SMN}
-\bar{R}^{R}{}_{SMN}
=
\bar{\nabla}_M \delta \hat{\Gamma}_{NS}^{R}
-
\bar{\nabla}_N \delta \hat{\Gamma}_{MS}^{R}.
\end{align}
The corresponding variations of the Ricci tensor
\(\hat R_{MN}\equiv \hat R^{R}{}_{MRN}\) and the Ricci scalar
\(\hat R\equiv \hat g^{MN}\hat R_{MN}\) then follow straightforwardly.

Combining these formulae with the conformal transformation
\eqref{E_tensor}, we obtain the linear perturbation of the Einstein tensor
of the physical metric \(g_{MN}=e^{-A}\hat g_{MN}\):
\begin{align}
\nonumber
\delta G_{MN}
=&\;
\frac{1}{2}
\left(
\bar{\nabla}_P \bar{\nabla}_M h_N{}^P
+
\bar{\nabla}_P \bar{\nabla}_N h_M{}^P
-
\bar{\Box} h_{MN}
-
\bar{\nabla}_M \bar{\nabla}_N h
\right)
\\
\nonumber
&-
\frac{1}{2}\bar g_{MN}
\left[
\bar{\nabla}_P\bar{\nabla}_Q h^{PQ}
-
\bar{\Box}h
-
h^{PQ}\bar R_{PQ}
\right]
-
\frac{1}{2}h_{MN}\bar R
\\
\nonumber
&-
\frac{3}{4}
\left(
\bar{\nabla}_M h_N{}^P
+
\bar{\nabla}_N h_M{}^P
-
\bar{\nabla}^P h_{MN}
\right)
\bar{\nabla}_P A
\\
\nonumber
&-
\frac{3}{2}h_{MN}\bar{\Box}A
-
\frac{3}{2}\bar g_{MN}
\left(
-\bar{\nabla}_P h^{PQ}\bar{\nabla}_Q A
+
\frac{1}{2}\bar{\nabla}^P h\,\bar{\nabla}_P A
-
h^{PQ}\bar{\nabla}_P\bar{\nabla}_Q A
\right)
\\
&+
\frac{3}{4}
\left(
h_{MN}\bar{\nabla}_P A\bar{\nabla}^P A
-
\bar g_{MN}h^{PQ}\bar{\nabla}_P A\bar{\nabla}_Q A
\right).
\label{delta_G_general}
\end{align}
Here
\begin{align}
h\equiv \bar g^{MN}h_{MN},
\qquad
\bar{\Box}\equiv \bar g^{MN}\bar{\nabla}_M\bar{\nabla}_N .
\end{align}

On the other hand, the source term also varies at linear order.  In the
tensor sector considered here, we do not include independent matter
perturbations.  The variation of the source term is therefore obtained from the
metric dependence of the background source.  With the convention that mixed
indices are raised by the background metric \(\bar g_{MN}\), namely
\(\bar{\mathcal T}^{P}{}_{N}\equiv
\bar g^{PQ}\bar{\mathcal T}_{QN}\), we have
\begin{align}
\delta \mathcal{T}_{MN}
=
\frac{1}{2}
\left(
h_{MP}\bar{\mathcal{T}}^{P}{}_{N}
+
h_{NP}\bar{\mathcal{T}}^{P}{}_{M}
\right).
\end{align}
Using the background Einstein equation together with the conformal
decomposition~\eqref{E_tensor}, this can be written as
\begin{align}
\nonumber
\delta \mathcal{T}_{MN}
=&\;
\frac{1}{2}
\left(
h_{MP}\bar{G}^{P}{}_{N}
+
h_{NP}\bar{G}^{P}{}_{M}
\right)+
\frac{3}{4}
h_{MP}
\left(
\frac{1}{2}\bar{\nabla}^{P}A\,\bar{\nabla}_{N}A
+
\bar{\nabla}^{P}\bar{\nabla}_{N}A
\right)
\\
&+
\frac{3}{4}
h_{NP}
\left(
\frac{1}{2}\bar{\nabla}^{P}A\,\bar{\nabla}_{M}A
+
\bar{\nabla}^{P}\bar{\nabla}_{M}A
\right)+
\frac{3}{2}
h_{MN}
\left(
-\bar{\Box}A
+
\frac{1}{2}\bar{\nabla}_{P}A\,\bar{\nabla}^{P}A
\right).
\label{per_T_MN}
\end{align}

Therefore, the linearized Einstein equation is obtained by equating
Eqs.~\eqref{delta_G_general} and~\eqref{per_T_MN}.  Although several terms
cancel already at this stage, the expression is still rather general.  We now
specialize it to the transverse-traceless tensor sector on the warped
cosmological background.

We now take the FLRW background~\eqref{bar_metric_FRW} and restrict the
fluctuation to the spatial tensor components \(h_{ij}\), setting all other
components to zero; see Eq.~\eqref{h_ij_2}.  We also impose the
transverse-traceless conditions~\eqref{TT_2}.  Under these assumptions, one can
show that
\begin{align}
h=\bar{g}^{MN}h_{MN}=0,
\qquad
\bar{\nabla}^{M}h_{MN}=0 .
\end{align}
Using these relations in the linearized Einstein equation
\(\delta G_{MN}=\delta\mathcal{T}_{MN}\), many terms drop and we obtain
\begin{align}
\bar{\Box}h_{MN}
+
\left(
\bar{R}^{S}{}_{MRN}
+
\bar{R}^{S}{}_{NRM}
\right)
h_{S}{}^{R}
-
\bar{g}_{MN}h^{RS}\bar{R}_{RS}
-
\frac{3}{2}
\bar{\nabla}^{R}h_{MN}\bar{\nabla}_{R}A
=0 .
\label{tensor_eom_general}
\end{align}
This equation reduces to the flat-brane result of Ref.~\cite{Csaki:2000fc} in
the limit \(a(t)=1\).

Let us now evaluate Eq.~\eqref{tensor_eom_general} for the spatial components
\((M,N)=(i,j)\).  For an exact de Sitter background, \(\dot H=0\), we find
\begin{align}
&\bar{\Box}h_{ij}
=
\underbrace{
-\ddot h_{ij}
+
H\dot h_{ij}
+
a^{-2}\partial^{2}h_{ij}
+
4H^{2}h_{ij}
}_{=\ \Box_{\rm dS}h_{ij}}
+
h_{ij}^{\prime\prime},
\\
&\left(
\bar{R}^{S}{}_{iRj}
+
\bar{R}^{S}{}_{jRi}
\right)
h_{S}{}^{R}
=
-2H^{2}h_{ij},
\\
&\bar{g}_{ij}h^{RS}\bar{R}_{RS}
=0,
\\
&\bar{\nabla}^{R}h_{ij}\bar{\nabla}_{R}A
=
A^{\prime}h_{ij}^{\prime}.
\end{align}
Here a dot denotes a derivative with respect to the cosmic time \(t\), while a
prime denotes a derivative with respect to the extra-dimensional coordinate
\(z\).  Therefore, Eq.~\eqref{tensor_eom_general} becomes
\begin{align}
-\Box_{\rm dS}h_{ij}
+
2H^{2}h_{ij}
+
\frac{3}{2}A^{\prime}h_{ij}^{\prime}
-
h_{ij}^{\prime\prime}
=
0,
\end{align}
which reproduces Eq.~\eqref{gravi_eom_RS2}.

\end{appendix}

\bibliographystyle{JHEP}
\bibliography{Refs}

\end{document}